\documentclass[12pt]{article}
\usepackage{amsmath,amssymb,color}
\usepackage{cite,epsf} 
\usepackage{pstricks}
\usepackage{color}
\usepackage{hyperref}
\usepackage{tikz}     
\usepackage{graphicx,caption}
\usepackage{subfigure}
\usetikzlibrary{
decorations.pathmorphing %library needed for e.g. decoration=snake
}

\usepackage{graphicx,array} 
\newcommand{\be}{\begin{equation}}
\newcommand{\ee}{\end{equation}}
\newcommand{\beq}{\begin{equation}}
\newcommand{\eeq}{\end{equation}}
\newcommand{\bea}{\begin{eqnarray}}
\newcommand{\eea}{\end{eqnarray}}

\newcommand{\ba}{\begin{eqnarray}}
\newcommand{\ea}{\end{eqnarray}}

\def\sin{\mbox{sin}}
\def\cos{\mbox{cos}}
\def\cot{\mbox{cot}}
\def\log{\mbox{log}}
\def\arctan{\mbox{arctan}}

\begin{document}

\begin{titlepage}
\vspace{10pt}
\hfill
{\large\bf HU-EP-20/29}
\vspace{20mm}
\begin{center}

{\Large\bf   Wilson loops for triangular contours\\[3mm]
with circular edges
}

\vspace{45pt}

{\large Harald Dorn 
{\footnote{dorn@physik.hu-berlin.de
 }}}
\\[15mm]
{\it\ Institut f\"ur Physik und IRIS Adlershof, 
Humboldt-Universit\"at zu Berlin,}\\
{\it Zum Gro{\ss}en Windkanal 6, D-12489 Berlin, Germany}\\[4mm]

\vspace{20pt}

\end{center}
\vspace{10pt}
\vspace{40pt}

\centerline{{\bf{Abstract}}}
\vspace*{5mm}
\noindent
We calculate Wilson loops in lowest order of perturbation theory for triangular contours whose edges are circular arcs. Based on a suitable disentanglement
of the relations between metrical and conformal parameters of the contours, the result fits perfectly in the structure 
predicted by the anomalous conformal Ward identity. The conformal remainder function depends in the generic 4D case on three cusp and on three torsion
angles. The restrictions on these angles imposed by the closing of the contour are discussed in detail and also for cases in 3D and 2D.

\vspace*{4mm}
\noindent

\vspace*{5mm}
\noindent
   
\end{titlepage}
\newpage

%\tableofcontents \newpage

\section{Introduction}
 %%%%%%%%%%%%%%%%%%%%%%%
%%%%%%%%%%%%%%%%%%%%%%%%%%%%%%%
In a recent paper \cite{Dorn:2020meb} we have derived anomalous conformal Ward identities for Wilson loops  along polygon-like contours,  whose edges are made of circular arcs. They enforce a factorised structure, with one factor depending on the distances between the corners and the cusp anomalous dimensions, just in the same way
as in correlation functions of local conformal operators with their conformal dimensions. The second factor is a remainder, depending only on  conformal invariant parameters
characterising the circular polygon. These parameters are the cusp angles, torsion angles and for more than three corners the usual cross ratios of the corner points.

Certainly it would have been useful to illustrate and underpin this general result by an explicit calculation. But it turned out to be a bit tricky
to disentangle the relevant equations between the involved metrical and conformal parameters in short time and to extract the finite piece remaining after subtraction of the UV divergent cusp terms.

The aim of this paper is to fill this gap by a lowest order calculation of the remainder function for the triangle with circular edges in Euclidean  ${\cal N}=4$ SYM. This we will do for the generic case which winds in full 4D space and comment also on triangles in 3D as well as on the planar situation.  

As a side effect, thereby we generate a rare example of a Wilson loop calculation for a contour not restricted to a subspace. The only other examples seem to be the Wilson loops for light-like polygons studied a lot in connection with the duality to scattering amplitudes \cite{Alday:2007hr},\cite{Drummond:2007aua}. But there the contours are fixed
by their corners and one has no parameters for some additional freedom of the edges. Wilson loops for toroidal contours winding in 3D have been studied in \cite{Drukker:2007qr}.

The Maldacena-Wilson loops for planar circular triangles have been studied non-perturbatively in \cite{Cavaglia:2018lxi}, but in a special limit which involves
large imaginary angles for the coupling of the scalars. For the treatment at strong coupling  via AdS/CFT of  generic smooth contours winding in full 4D  
the local conformal characteristics of the contour have been related to the boundary data of the Pohlmeyer fields in \cite{He:2017cwd}.

Wilson loops for triangles with straight edges have been used  in QCD for applications to baryon phenomenology, see e.g. \cite{Bali:2000gf,Kuzmenko:2002zs,Andreev:2015riv} and references therein. There is also recent work on the large size behaviour for standard planar polygons with straight edges \cite{Pobylitsa:2019ewu}.\\

\begin{figure}[h!]
\begin{center}
 \includegraphics[width=8cm]{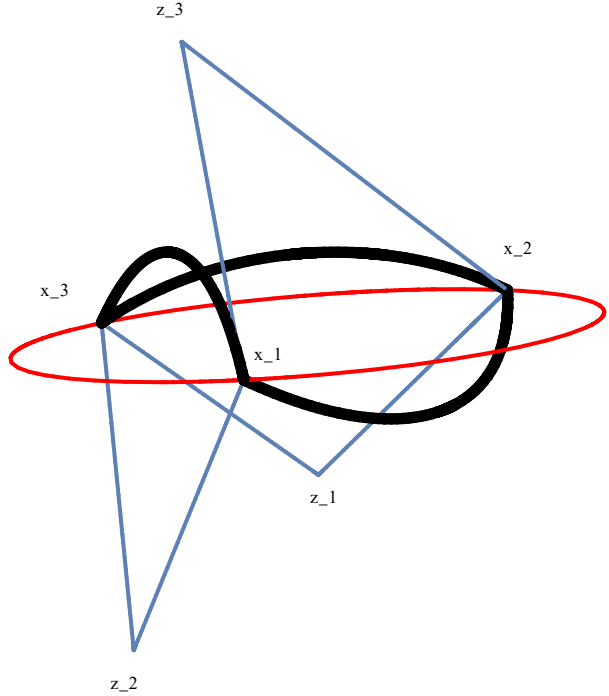}
\end{center}
\caption {\it A triangle with corners $X_1,X_2,X_3$ and circular edges is shown in black. The corresponding circum circle is depicted in red. The blue lines are the radii connecting the corners with the centers related to the  circular edges.  }
\label{fig:triangle}
\end{figure}

For a first visualisation of the contour under consideration we show fig.\ref{fig:triangle}, taken from  \cite{Dorn:2020meb}. For obvious reasons it is a 3D object.
In the generic 4D case one has eleven related independent metrical invariants  \cite{Dorn:2020meb}. As those can be chosen the three distances between the corners $X_j$
\beq
D_{ij}~=~\vert X_i-X_j\vert~,
\eeq
the radii $R_j$ of the three edges, the three distances between the centers of circular edges and two of the distances between a corner $X_j$ and the center of the opposite
circular edge $Z_j$. Furthermore, there are only six conformal invariants, three angles $\alpha_j$ and three angles $\beta_j$, where $\alpha_j$ is the cusp angle at corner $X_j$ and
$\beta_j$ the angle between the circumcircle and the edge number $j$. Due to their meaning in our geometrical setting, all angles $\alpha_j$ and $\beta_j$ are a priori restricted to the interval $(0,\pi)$. In the planar case the $\beta_j$  are fixed by the cusp angles, but as soon as they vary independently
from these constraints, the contour winds out of the plane fixed by the corners.\footnote{For that reason we call them torsion angles.} Our task will be to find a representation of the Wilson loop in terms of the set  $(D_{ij},\alpha_j,\beta_j)$.
%%%%%%%%%%%%%%%%%%%%%%%%%%%%%%%%%%%%%%%%%%%%
%%%%%%%%%%%%%%%%%%%%%%%%%%%%%%%%%%%%%%%%%%%%%%%

The Maldacena-Wilson loop is defined by
\beq
{\cal W}~=~\big \langle \frac{1}{N}\mbox{tr}\,\mbox{P\,exp}\int(i A_{\mu}\dot x^{\mu}+\vert\dot x\vert \phi_i\theta^i)dt~\big\rangle\label{MW}~.
\eeq
Let us consider here only the case with constant $\theta^i$ along the whole contour. In lowest order we have \footnote{$C_F=\frac{N^2-1}{2N}$ for $SU(N)$ gauge group.}
\beq
\log\,{\cal W}~=~\frac{g^2C_F}{4\pi^2}~I~+~{\cal O}(g^4)~,\label{W-I}
\eeq
\beq
I~=~I^{\mbox{\scriptsize scalar}}~-~I^{\mbox{\scriptsize vector}}~=
~\frac{1}{2}\,\int\int\frac{\vert \dot x_1\vert \vert \dot x_2\vert -\dot x_1\dot x_2}{(x_1-x_2)^2+a^2}\,dt_1\,dt_2~,\label{I}
\eeq
understood as limit $a\rightarrow 0$ after subtraction of the divergent pieces. 

This regularisation has been used in the first paper on the cusp anomalous dimension for the pure gauge field case \cite{Polyakov:1980ca} and also in the first
paper on the supersymmetric case \cite{Drukker:1999zq}. Later calculations of higher orders have mainly used dimensional regularisation.
Note that for that purpose they could use straight edges. Our choice seems technically more convenient
for the evaluation of the finite piece for a curved contour.

According to \cite{Dorn:2020meb} the Wilson loop for a triangle with circular edges as in fig.\ref{fig:triangle} has the following structure  \footnote{As in \cite{Dorn:2020meb} we neglect an overall factor $\mu^{-\sum \Gamma_j}$, depending on the RG scale $\mu$.}
\beq 
{\cal W}~=~D_{12}^{\Gamma_3-\Gamma_1-\Gamma_2}D_{23}^{\Gamma_1-\Gamma_2-\Gamma_3}D_{13}^{\Gamma_2-\Gamma_1-\Gamma_3}~\Omega(\alpha_j,\beta_j)~.\label{ward}
\eeq
$\Gamma_j=\Gamma(\alpha_j)$ are the cusp anomalous dimensions. They are  \footnote{$\alpha$ chosen as the opening angle of a cusp, i.e. $\alpha=\pi$ as smooth case.}
\beq
\Gamma(\alpha)~=~\frac{g^2C_F}{4\pi^2}~\gamma(\alpha)~+~{\cal O}(g^4) ~,\label{Gamma}
\eeq
with
\beq
\gamma(\alpha)=-\,(\pi-\alpha)~\frac{1+\cos\alpha}{\sin\alpha}~~~~~\mbox{or}~~~~~\gamma(\alpha)=-\,(1+(\pi-\alpha)\cot\alpha )~\label{gamma}
\eeq
for the Maldacena-Wilson loop \eqref{MW} with smooth contour in the internal space or the Wilson loop for pure gauge fields, respectively.

The remainder function $\Omega$ depends on three, five, or six independent conformal invariants in $D= 2,~D=3,$ or $D=4$ dimensions, respectively. As these conformal invariants can be chosen the 3 cusp angles $\alpha_j$ in $D=2$, due to a closing condition five angles out of the three cusp angles and the three torsion angles $\beta_j$ in  $D=3$ and all
six angles without a local constraint in  $D=4$. 

In calculating \eqref{I} one has to add the three pieces where $x_1(t_1)$ and $x_2(t_2)$ are on the same edge and the three pieces where they are on the adjacent edges of one of the corner points, i.e.
\beq
I~=~\sum_{j=1}^3(E_j~+~C_j)~.\label{IEC}
\eeq
Only after taking into account the condition that  the single terms in \eqref{IEC} combine to a closed contour, one can organise the dependence on metrical and conformal invariants in a manner required to fit \eqref{ward}. 

To illustrate this issue, and as a warm up, let us consider in the next section the limiting case of a standard triangle
with straight edges (then we have only 2 conformal invariants, since $\sum \alpha_j=\pi$).

In the generic case there are more metrical invariants beyond the $D_{ij}$'s, and the relations between them and the conformal invariants are far more
involved. Therefore, we use in section 3 conformal invariance to calculate the corner building blocks $C_j$ in a suitable conformal frame. This allows
from the beginning a parameterisation with maximal use of conformal invariants.\footnote{For circular triangles in a plane there has been used in \cite{Cavaglia:2018lxi}
a clever parameterisation of the original geometry in terms of the corner points and cusp angles, without using the radii . We do not know a suitable extension beyond the planar case.}

Several technical calculations are sketched in appendices. 
%%%%%%%%%%%%%%%%%%%%%%%%%%%%% 
%%%%%%%%%%%%%%%%%%%%%%%
\section{The standard triangle with straight edges}
Due to the straight edges, in this case the scalar and vector contributions are trivially related
\beq
C_j^{\mbox{\scriptsize vector}}~=~-\,\cos\alpha_j~C_j^{\mbox{\scriptsize scalar}}~,~~~~~~~
E_j^{\mbox{\scriptsize vector}}~=~E_j^{\mbox{\scriptsize scalar}}~.\label{scalarvector}
\eeq
For the corner term, e.g. for the corner at $X_3$ we have
\beq
C_3^{\mbox{\scriptsize scalar}}~=~\int_0^{D_{23}}dt_1\int_0^{D_{13}}dt_2\,\frac{1}{t_1^2+t_2^2-2t_1t_2\,\cos\alpha_3+a^2}~.
\eeq
Performing the $t_2$-integration we get
\beq
C_3^{\mbox{\scriptsize scalar}}=\int_0^{D_{23}}\frac{d\,t}{\sqrt{a^2+t^2\sin^2\alpha_3}}
\Big (\arctan\big (\frac{D_{13}-t\,\cos\alpha_3}{\sqrt{a^2+t^2\sin^2\alpha_3}}\big )+\arctan\big (\frac{t\,\cos\alpha_3}{\sqrt{a^2+t^2\sin^2\alpha_3}}\big )\Big ).
\eeq
To evaluate this integral in the limit of vanishing UV regulator $a$, we split into pieces and make suitable subtractions\footnote{Some details for the similar evaluation of the integrals in the generic case of a triangle with circular edges are presented in appendix B.} to arrive at ($\Theta(x)$ denotes the UnitStep  function)
\bea
C_3^{\mbox{\scriptsize scalar}}&=&\frac{\pi-\alpha_3}{\sin\alpha_3}\,\log\frac{2D_{23}\,\sin\alpha_3}{a }~+~\frac{\alpha_3-\frac{\pi}{2}}{\sin\alpha_3}\,\log\big (\sin\alpha_3+\sqrt{1+\sin^2\alpha_3}\big )\nonumber\\[2mm]
&&+\int_0^{\infty}\frac{d\,t}{\sqrt{1+t^2\sin^2\alpha_3}}
\Big ( \arctan\big (\frac{t\,\cos\alpha_3}{1+t^2\sin^2\alpha_3}\big )+(\alpha_3-\frac{\pi}{2})\,\Theta(t-1)\Big )\nonumber\\[2mm]
&&+~\frac{1}{\sin\alpha_3}~\mbox{Im Li}_2\big (\frac{D_{23}}{D_{13}}\,e^{-i\,\alpha_3}\big )~+~{\cal O}(a\,\log a)~.
\eea
Although not obviously, this is symmetrically in $D_{13}\leftrightarrow D_{23}$ as can be checked by using appropriate functional relations of the dilogarithm.

The edge contribution, e.g. for $E_3^{\mbox{\scriptsize scalar}}$ is
\bea
E_3^{\mbox{\scriptsize scalar}}&=&\frac{1}{2}\,\int_0^1d\,t_1\int_0^1d\,t_2~\frac{1}{(t_1-t_2)^2+\frac{a^2}{D_{12}^2}}\nonumber\\
&=&\frac{\pi\,D_{12}}{2\,a}~+~\log\frac{a}{D_{12}}~-~1~+~{\cal O}(a^2)~.
\eea
As mentioned above, the building blocks for \eqref{IEC} depend on both conformal and pure metrical invariants. 
%%%%%%%%%%%%%%%%%%%%%%%%%%%%,%%i
 %%%%%%%%%%%%%%%%%%%%%%%%%%%%%%%%%%%%%%%%%%%%
\begin{figure}[h]
\begin{minipage}{0.45\textwidth}
\includegraphics[width=\textwidth]{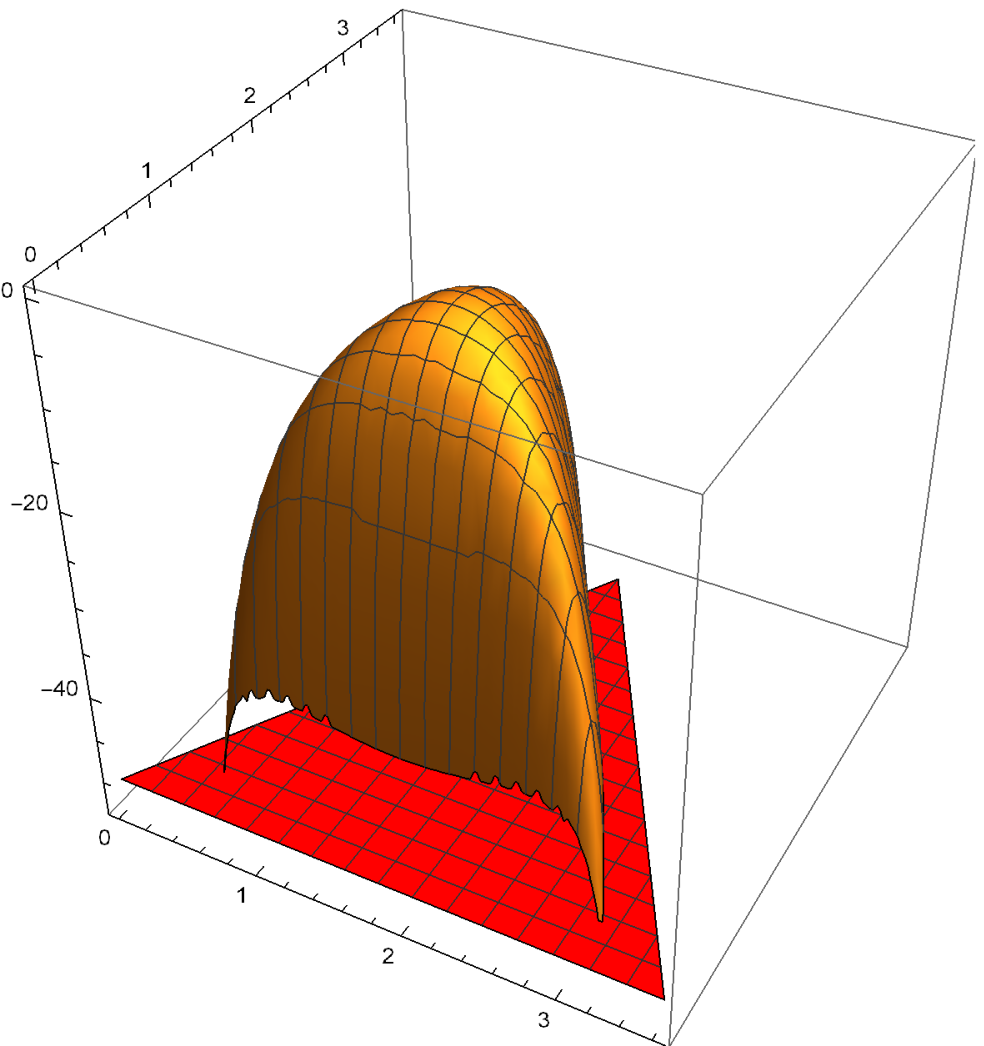}
\end{minipage}
\begin{minipage}{0.45\textwidth}
\includegraphics[width=\textwidth]{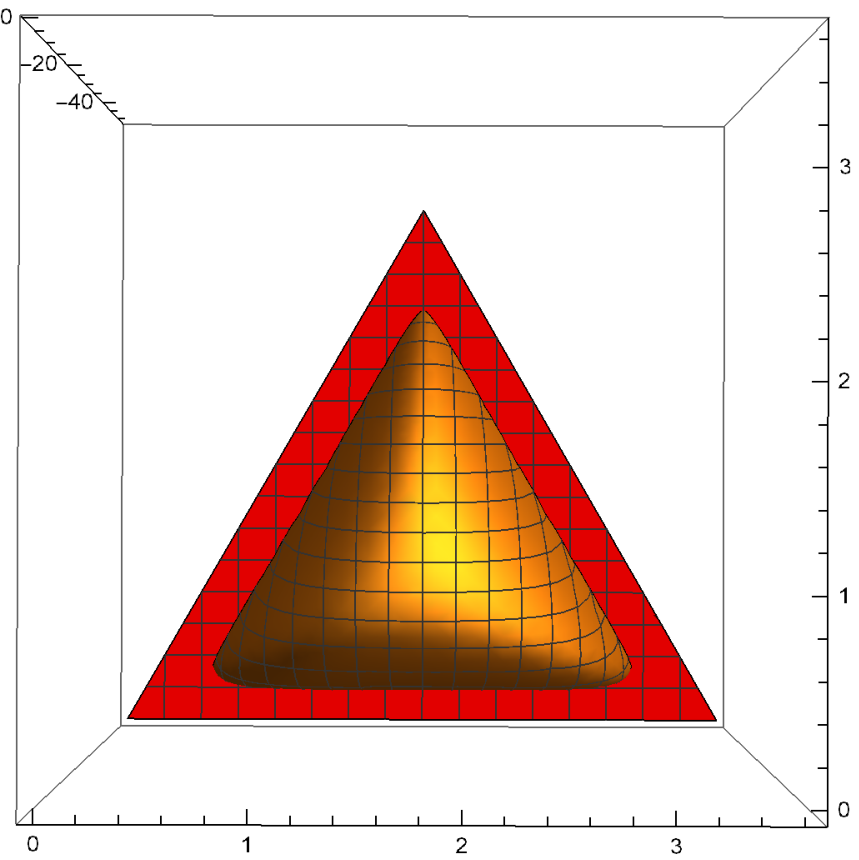}
\end{minipage}
\caption {\it The remainder $\omega^{\mbox{\scriptsize st}}(\alpha_1,\alpha_2,\alpha_3)$ from two different perspectives. The angles are given by the distance to the edges of the red triangle in the $(x,y)$-plane. The plots exclude angles smaller than $0.3$. }
\label{standard-triangle}
\end{figure}
%%%%%%%%%%%%%%%%%%%%%%%%%%%%%%%%%%%
Now we implement the closing to a triangle. Then, due to the sine law for standard triangles, we can replace the quotients of distances in the arguments
of the dilogarithms by quotients of sines of the opposite angles. Then the only remaining dependence of the renormalised $I$ from \eqref{IEC} on the distances $D_{ij}$ has the form $-\gamma(\alpha_3)\log D_{23}-\gamma(\alpha_2)\log D_{12}-\gamma(\alpha_1)\log D_{13}$. To fit this to the structure required by \eqref{ward}, we add e.g. to the factor multiplying $-\gamma(\alpha_3)$ the term $\log \frac{D_{13}}{D_{12}}$ and subtract it again after
using the triangles sine law. This yields  finally for the standard triangle remainder function of the Maldacena-Wilson loop with \eqref{scalarvector},\eqref{IEC},\eqref{gamma},\eqref{ward},\eqref{W-I}
\bea
\log \,\Omega^{\mbox{\scriptsize st }}&=&\frac{g^2C_F}{4\pi^2}~\omega^{\mbox{\scriptsize st }}~+~{\cal O}(g^4)~,\nonumber\\
\omega^{\mbox{\scriptsize st }}&=&\sum_{j=1}^3\left ( \gamma(\alpha_j)\,\log\,\frac{\sin\,\alpha_{j-1}}{2\,\sin^2\alpha_j}~+~\frac{1+\cos\,\alpha_j}{\sin\,\alpha_j}\,\mbox{Im Li}_2\big(\frac{\sin\,\alpha_{j+1}}{\sin\,\alpha_{j-1}}e^{-i\alpha_j}\big)\right . \nonumber\\
&&+\left . \gamma(\alpha_j)\,\frac{\alpha_j -\frac{\pi}{2}}{\alpha_j-\pi}\,\log\big (\sin\,\alpha_j+\sqrt{1+\sin^2\alpha_j}\big )\right .\\
&&+\left . \int_0^{\infty}dt\frac{1+\cos\,\alpha_j }{\sqrt{1+t^2\sin^2\alpha_j}}
\Big ( \arctan\big (\frac{t\,\cos\,\alpha_j}{\sqrt{1+t^2\sin^2\alpha_j}}\big )+\big (\alpha_j-\frac{\pi}{2}\big )\,\Theta(t-1)\Big )\right ),\nonumber
\eea
with the constraint $\sum_j\alpha_j=\pi$.\footnote{Here and below we use the identification $j\pm 3=j\,\mbox{mod}\,3$.}

To plot a numerical evaluation based on this formula in a manner symmetric in the $\alpha_j$'s, we use the fact that for each point inside  an equilateral triangle the sum 
of its distances to the edges is constant. For an edge length $2\pi/\sqrt{3}$ this sum is just equal to $\pi$.
 Then putting this equilateral triangle in the $(x,y)$-plane we have 
\beq
\alpha_1=y~,~~~\alpha_2=\frac{2\pi-\sqrt{3}\,x-y}{2}~,~~~\alpha_3=\frac{\sqrt{3}\,x-y}{2}~.
\eeq
The remainder  $\omega^{\mbox{\scriptsize st}}(\alpha_1,\alpha_2,\alpha_3)$, via this formula as a function of $x$ and $y$, is presented in fig.\ref{standard-triangle}.
Its maximal value is $-1.05373$, obtained for equal angles. Since $\omega^{\mbox{\scriptsize st}}$ diverges if one of the angles approaches zero, a certain  neighbourhood of zero is excluded in the plots. 
%%%%%%%%%%%%%%%%%%%%%%%%%%%%,%%i
%%%%%%%%%%%%%%%%%%%%%%%%%%%%%%%%
\section{The generic triangle with circular edges}
Denoting by $\delta_j$ the opening angle of the circular edge number $j$, i.e.
\beq
\sin\frac{\delta_j}{2}~=~\frac{D_{j-1,j+1}}{2R_j}~,\label{delta-j}
\eeq
we get by straightforward evaluation
\beq
E_j^{\mbox{\,\scriptsize scalar}}~=~\frac{\pi R_j\delta_j}{2a}~+~\log\frac{a}{D_{j-1,j+1}}~-~1~+~{\cal O}(a^2\log a)
\eeq
and
\beq
E_j^{\,\mbox{\scriptsize vector}}~=~E_j^{\mbox{\,\scriptsize scalar}}~-~\frac{\delta_j^2}{4}~+~{\cal O}(a^2)~.
\eeq
Therefore the edge contribution to the supersymmetric Maldacena-Wilson loop is
\beq
\sum_j(E_j^{\mbox{\,\scriptsize scalar}}-E_j^{\,\mbox{\scriptsize vector}})~=~\frac{1}{4}\sum_j\delta_j^2~+~{\cal O}(a^2\log a)~.\label{sumE}
\eeq
To evaluate the corner term $C_3$, we map the corner number 3 with its two adjacent edges to a conformal frame where the  corner is at infinity, see appendix A and \cite{Dorn:2020meb}. The other corner terms are then given by cyclic permutations of the indices. 

Let us start with the scalar contribution. Without regularisation $(a=0)$ the integrand of the scalar contribution to \eqref{I} is invariant under all
conformal transformations.  For $a>0$ we get, after a shift to move corner 3 to the origin followed by an inversion and a dilatation to scale the distance between
corners 1 and 2 to one, 
\beq
\frac{\vert\dot x_1(t_1)\vert \vert\dot x_2(t_2)\vert}{(x_1-x_2)^2+a^2}~=~\frac{\vert\dot y_1(t_1)\vert \vert\dot y_2(t_2)\vert}{(y_1-y_2)^2+b_3^2y_1^2y_2^2}~.\label{x-y}
\eeq
The abbreviation $b_3$ is defined by
\beq
b_3~=~aD_3~, ~~~~~~~D_3~=~\frac{D_{12}}{D_{13}D_{23}}~.\label{b3}
\eeq
Proceeding with 
\beq
y_1(t_1)~=~Y_2~+~t_1\,e_2~,~~~~y_2(t_2)~=~Y_1~+~t_2\,e_1~,\label{y-conf-frame}
\eeq
and  $Y_j$ and $e_j$  defined as in appendix A,  \footnote{The use of this special form still requires some rotations, but these do not change the form of the r.h.s. of \eqref{x-y}.}
we get with \eqref{beta-theta-phi-1},\eqref{beta-theta-phi-2},\eqref{alpha-theta-phi-3}
\beq
(y_1(t_1)-y_2(t_2))^2~=~1+t_1^2+t_2^2-2t_1t_2\,\cos\alpha_3+2t_1\,\cos\beta_1+2t_2\,\cos\beta_2~.
\eeq
Taking the results of appendix B ($b\rightarrow b_3,~\alpha\rightarrow \alpha_3$) we obtain from \eqref{b3}, \eqref{app-I},\eqref{J-final},\eqref{Q}
\beq
C_3^{\mbox{\scriptsize scalar}}~=~\frac{\pi-\alpha_3}{\sin\alpha_3}\,\log\frac{D_{13}D_{23}}{D_{12}}~+~\frac{\pi-\alpha_3}{\sin\alpha_3}\,\log(2\sin\alpha_3)~+~Q(\alpha_3,\beta_1,\beta_2)~.\label{C3scal}
\eeq

Now we turn to the more involved issue of the vector contribution . If the analogue of \eqref{x-y} would be true, with the nominators replaced by the scalar products of the tangents, $C_3^{\mbox{\scriptsize vector}}$ would be given by $C_3^{\mbox{\scriptsize scalar}}$ multiplied by $-\cos\alpha_3$. However,  due to the violation of the invariance of even the unregularised integrand we get an additional contribution
\beq
C_3^{\mbox{\scriptsize vector}}~=~-\cos\alpha_3\,C_3^{\mbox{\scriptsize scalar}}~-~A_3~.\label{C3}
\eeq
Using a representation of the invariance breaking term as in \cite{Drukker:2000rr} we calculate the  $A_j$ in appendix C and get from \eqref{sum-A}
\beq
\sum_j A_j~=~\pi^2+\sum_j\big (\beta^2_j~-~(\pi-\alpha_j)^2~-~\frac{1}{4}\delta_j^2\big )~.\label{sumA}
\eeq
Collecting now \eqref{sumE},\eqref{C3scal},\eqref{C3},\eqref{sumA} for use in \eqref{ward} and \eqref{IEC}, we get with \eqref{Gamma},\eqref{gamma} for the Maldacena-Wilson loop ($Q$ as defined in appendix B, \eqref{Q}. For the planar case see also appendix D, \eqref{Qsame},\eqref{Qoppo}.)
\beq
\log\, \Omega ~=~\frac{g^2C_F}{4\pi^2}~\omega~+~{\cal O}(g^4)~,\label{omega-final}
\eeq
\beq
\omega=\pi^2+\sum_j\big ( (1+\cos\alpha_j)\,Q(\alpha_j,\beta_{j-1},\beta_{j+1})+\beta_j^2-(\pi-\alpha_j)^2-\gamma(\alpha_j)\log(2\,\sin\alpha_j)\big)~.\nonumber 
\eeq
Furthermore, we note that the terms depending on the metrical invariants $D_{ij}$ just organise to fit their appearance in \eqref{ward}. As expected, the dependence on the radii of the circular edges, encoded in the $\delta_j$, cancels in the final result.

In a similar manner we find that for the pure gauge Wilson loop $\omega$ is replaced by
\bea
\omega^{\mbox{\scriptsize vec}}&=&\pi^2+3\label{omega-vector}\\
&+&\sum_j\big ( \cos\alpha_j\,Q(\alpha_j,\beta_{j-1},\beta_{j+1})+\beta_j^2-(\pi-\alpha_j)^2+(\pi-\alpha_j)\,\cot\alpha_j\log(2\,\sin\alpha_j)\big)~.\nonumber
\eea

\begin{figure}[h]
\begin{minipage}{0.5\textwidth}
\includegraphics[width=\textwidth]{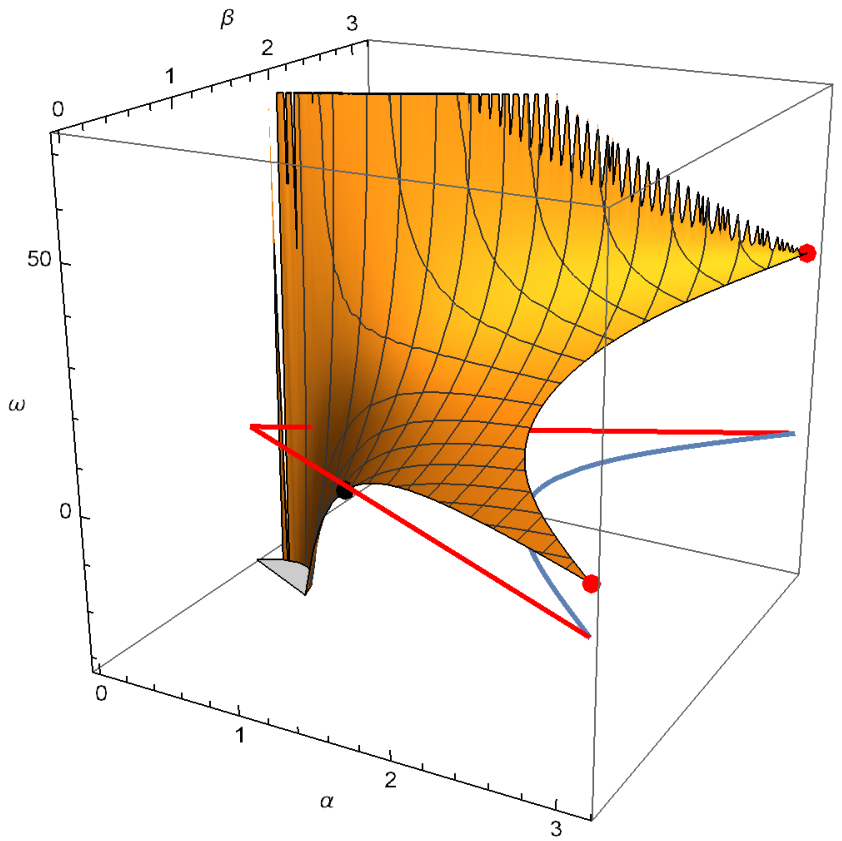}
\end{minipage}
\begin{minipage}{0.5\textwidth}
\includegraphics[width=\textwidth]{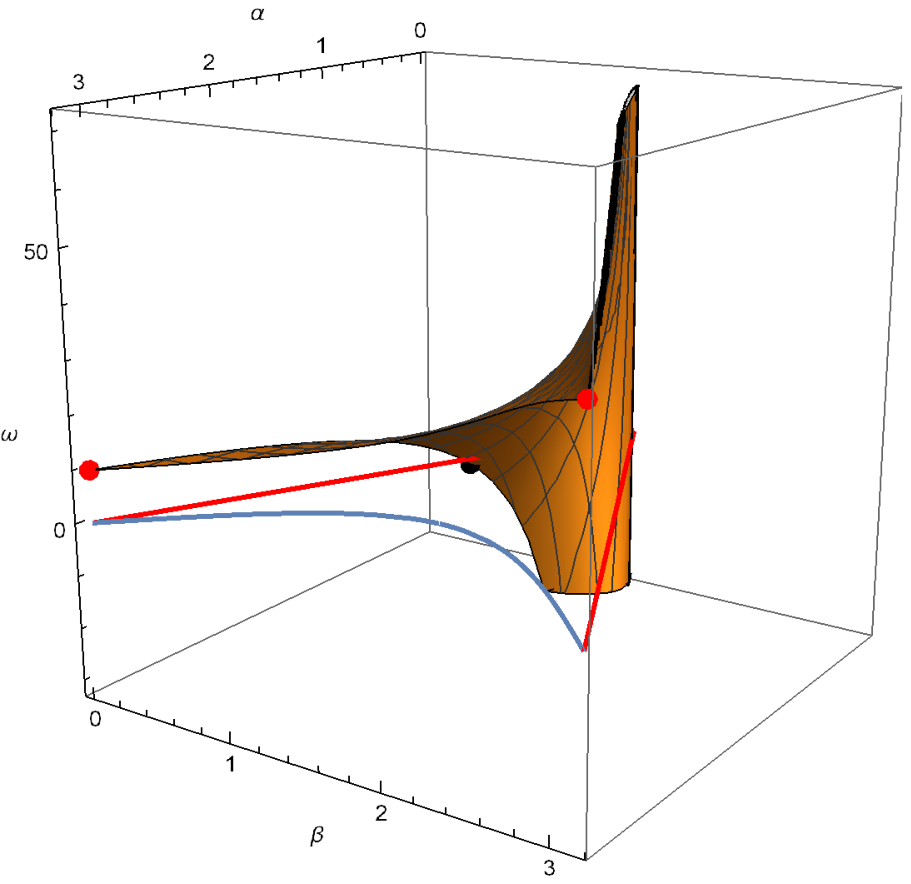}
\end{minipage}
\caption {\it The remainder $\omega(\alpha,\alpha,\alpha,\beta,\beta,\beta)$ from two different perspectives. The blue and red lines in the $(\alpha,\beta)$-plane indicate the boundary of the geometrically allowed angles. The red line corresponds to planar triangles and the blue line to those embedded in a three-dimensional subspace. The  black dot with $\alpha=\beta=\frac{\pi}{3}$ indicates the equilateral triangle with straight edges and the red dots mark  the circle and the twice traversed circle, respectively.}
\label{rem-alpha-beta}
\end{figure}

\begin{figure}[h!]
  \includegraphics[width=12cm]{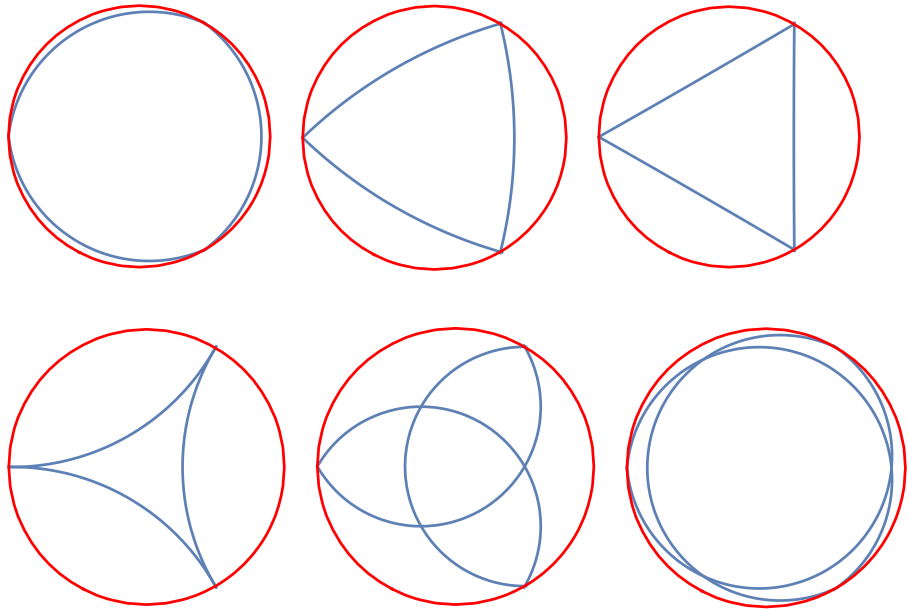} 
\centering
\caption{\it The planar situation varying along the red line in fig.\ref{rem-alpha-beta} from a contour close to the simple circle to one close to the twice traversed circle.The circumcircle is shown in red.} 
\label{plot-planar}
\end{figure}

To get some visual impression of the effect of torsion of the circular triangle, we look on the case, where all cusp angles $\alpha_j$ are equal and all torsion angles $\beta_j$
are equal. Then we can generate in fig.\ref{rem-alpha-beta} a 3D-plot of the remainder $\omega$ as a function of two variables $\alpha$ and $\beta$. The region in the $(\alpha,\beta)$-plane, allowed
by geometry in 4D, is fixed by the discussion in appendix A, \eqref{qj},\eqref{q-pm1},\eqref{closing-final} as follows
\beq
-\frac{1}{2}\, \leq\,q\leq\,1~~~~\mbox{with}~~~~q\,=\,\frac{\cos\alpha+\cos^2\beta}{\sin^2\beta}~.
\eeq
$q=1$ is reached for planar cases and $q=-\frac{1}{2}$ corresponds to cases, where the triangle is nonplanar, but can be embedded in a 3D subspace. The plot in fig.\ref{rem-alpha-beta} shows runaway behaviour in the vicinity of $(\alpha=0,\,\beta=\frac{\pi}{2})$ and along the part of the red line with $\beta\geq\frac{\pi}{2}$.
In  the first case this is due to the obvious singularity related to the degeneration of the cusps to spikes with zero opening angle, see the fourth picture in fig.\ref{plot-planar} (compare also  \cite{Dorn:2018als}). To understand the second case, we have
to recall, that $\beta$ measures the angle between the circular edges and the circumcircle. $q=1$ implies a planar situation with all edges on the same side
of the circumcircle. If then $\beta>\frac{\pi}{2}$, one necessarily has self-intersections of the Wilson loop, see the last two pictures in \nolinebreak fig.\ref{plot-planar} and the discussion at the end of appendix B.
%%%%%%%%%%%%%%%%%%%%%%%%%%%%%%%%%%%%%%%%%%%%%%%%%%%%%%%%%%%%%%%%%%5
%%%%%%%%%%%%%%%%%%%%%%%%%%%%%%%%%%%
%%%%%%%%%%%%%%
%%
%% Generated in  2020-09-30.nb , The analogon for the pure vector case generated in 2020-10-05.nb
%%
%%%%%%%%%%%%%%%%%%%%%%%%%%%%%%%%%%%%%%
%%%%%%%%%%%%%%%%%%%%%%%%%%%
%%%%%%%%%%%%%%%%%%%%%%%%%%%.
%%%%%%%%%%%%%%%%%%%%%%%%%%%%%%%%%%%%%%%%%%%%%%%%%%%%%%%%%%%%%%%%%%5
%%%%%%%%%%%%%%%%%%%%%%%%%%%%%%%%%%%
%%%%%%%%%%%%%%
%%
%% Generated in  2020-10-04.nb 
%%
%%%%%%%%%%%%%%%%%%%%%%%%%%%%%%%%%%%%%%
%%%%%%%%%%%%%%%%%%%%%%%%%%%
%%%%%%%%%%%%%%%%%%%%%%%%%%%.

Before closing this section, we still  mention some consistency checks. At first let us consider the case $\beta =\frac{\pi-\alpha}{2}$. In the limit $\alpha\rightarrow\pi$ the corresponding contour becomes a circle (just the circumcircle of the triangle setting), and for both the supersymmetric as well as the pure gauge loop we expect the well-known result $\pi^2$. Then the  function $Q$ diverges 
logarithmically. But due to its prefactor  the expected result is obvious for the case of \eqref{omega-final}. To get the same result for \eqref{omega-vector}
we can rely on
\beq
\lim_{\alpha\rightarrow\pi}\big (\cos\alpha\,Q(\alpha,\frac{\pi- \alpha}{2},\frac{\pi-\alpha}{2})+(\pi-\alpha)\,\cot\alpha\,\log(2\,\sin\alpha)\big )~=~-1~,\label{circle-limit}
\eeq
which has been checked numerically.

A similar limit is   $\alpha\rightarrow\pi$ with  $\beta =\frac{\pi+\alpha}{2}$. Then the contour approaches a twice traversed circle, see last picture in fig.\ref{plot-planar}.
Due to the degenerating self-intersections, the function $Q$ diverges faster than a logarithm. Nevertheless, due to its prefactor in \eqref{omega-final} one finds
for the supersymmetric remainder the expected value $4\pi^2$. But for the pure vector remainder there remains a divergence.\footnote{This is not a desaster, since limits of the geometrical settings not necessarily have to commute with the renormalisation procedure.}
%%%%%%%%%%%%%%%%%%%%
%%
%% Check in 2020-07-22.nb and 2020-10-05
%%
%%%%%%%%%%%%%%%%%%%%%%%

A third check comes from the limit, in which the generic circular triangle approaches a standard triangle with straight edges. First, as discussed in appendix A, the planar case, still with circular edges on the same side of the circumcircle, corresponds to $\alpha_j+\beta_{j-1}+\beta_{j+1}=\pi$, i.e.
\beq
\omega^{\mbox{\scriptsize planar}}\big (\alpha_j\big )~=~\omega\big (\alpha_j,\frac{\pi+\alpha_j-\alpha_{j-1}-\alpha_{j+1}}{2}\big )~.
\eeq
Then $\omega^{\mbox{\scriptsize planar}} $, calculated in this section by mapping parts of the contour to a conformal frame, has to be equal to $\omega^{\mbox{\scriptsize st}}$, calculated without any mapping in the previous section, as soon as the sum of cusp angles is equal to $\pi$. This we have checked numerically over the full range of angles obeying this constraint. 
%%%%%%%%%%%%%%%%%%%%%%%%%%%%%%%%%%%
%%%%%%%%%%%%%%
%%
%% Check in 2020-09-29.nb 
%%
%%%%%%%%%%%%%%%%%%%%%%%%%%%%%%%%%%%%%%
%%%%%%%%%%%%%%%%%%%%%%%%%%%
%%%%%%%%%%%%%%%%%%%%%%%%%%%.
 %%%%%%%%%%%%%%%%%%%%%%%%%%%%%%%%%%%%%%%%%%%%

A last comment concerns the generalisation to triangles where at the corners also a discontinuity in the coupling of the scalars is allowed. Let $\theta(t)$
in \eqref{MW} be constant and equal to $\theta _{(j)}$ along edge number $j$, but
\beq
\cos\, \chi_j~=~\theta_{(j-1)}\cdot\theta_{(j+1)}~.
\eeq
This has no effect on the edge contributions  and on the vector part of the corner terms in \eqref{IEC}.  Only  the scalar corner terms pick up factors
$\cos\chi_j$. This leads to a modification of the cusp anomalous dimension from \eqref{gamma}  to \cite{Drukker:1999zq}
\beq
\gamma(\alpha,\chi)=-\,(\pi-\alpha)~\frac{\cos\chi+\cos\alpha}{\sin\alpha}~.
\eeq
Then use of this new form of $\gamma$ has to be made in \eqref{omega-final}, and furthermore the prefactors $(1+\cos\alpha_j)$ of the $Q$-terms become $(\cos\chi_j+\cos\alpha_j)$.

The BPS case $\chi_j=\pi-\alpha_j$ simplifies to
\beq
\omega^{\mbox{\scriptsize BPS}}~=~\pi^2+\sum_j\big (\beta_j^2-(\pi-\alpha_j)^2\big)~.
\eeq
Furthermore, restricting ourselves to the planar case, and there e.g. to the same side situation with no selfcrossing $\alpha_j=\pi-\beta_{j-1}-\beta_{j+1}$, see \eqref{planar},
we get
\beq
\omega^{\mbox{\scriptsize BPS}}_{\mbox{\scriptsize planar}}~=~\frac{1}{4}\,\Big(\sum_j\alpha_j-\pi\Big )\Big (5\pi-\sum_j\alpha_j\Big )~.
\eeq
For all $\alpha_j$ equal to $\pi$ we have the limiting case of a circle and get again the well-known result $\pi^2$. The case $\sum_j \alpha_j=\pi$
covers standard triangles with straight edges and yields the result zero.

With the just discussed generalisation to cusps in the scalar coupling one can also specialise in a limit, where only scalar ladder diagrams survive 
\cite{Cavaglia:2018lxi},\\
 i.e. $g\rightarrow 0,~N\rightarrow\infty,~\chi_j\rightarrow i\;\infty$, with $ \hat g_j^2=\frac{g^2C_F}{4\pi^2}\,\cos\chi_j$ fixed. Then one gets
\beq
\log\,\Omega\vert_{CGL}~=~\sum_j\hat g^2_j\big( Q(\alpha_j,\beta_{j-1},\beta_{j+1})~+~\frac{\pi-\alpha_j}{\sin\alpha_j}\log (2\sin\alpha_j)\big)~+~{\cal O}(\hat g^4_j)~.
\eeq
For planar triangles in \cite{Cavaglia:2018lxi} they have summed all orders. Our last formula, specialised to the planar case, corresponds then to the $\hat g^2$ order of their summation result.

%%%%%%%%%%%%%%%%%%%%%%%%%%%%
\section{Summary and conclusions}
Our main result is the lowest order calculation of the remainder function of  Wilson loops  in ${\cal N}=4$ SYM for  triangles with circular edges,  both for the supersymmetric \eqref{omega-final} as well as for the pure vector case \eqref{omega-vector}. It is given as a function of only conformal invariant parameters by standard functions and a convergent one-dimensional integral, suitable for immediate numerical evaluation. For the planar case a full  representation in terms of standard functions has been found. In the limiting situations approaching a circle or a twice traversed circle we found agreement with the well-known results. Also the comparison of the  respective limit with an independent calculation for the triangle with straight edges  has given agreement.

We also commented on the obvious generalisation to cases where the coupling to the scalars is allowed to jump at the corners. In the related BPS case
the dependence on the distances between the corners is absent and the remainder functions becomes a quadratic expression in the cusp and torsion angles,
which in the planar situation simplifies even more to a quadratic expression in the sum of the cusp angles. It is zero for the standard triangles with straight
edges. While in 2D  these are the only cases with vanishing radiative corrections, in higher dimensions there is a whole variety with circular edges.

The appendix A extends the analysis of the conformal geometry of circular triangles started in \cite{Dorn:2020meb}. We introduced  for each building block,
consisting out of a corner and its two adjacent edges, an off-planarity parameter $q_j$. This allowed a formulation of  the constraints imposed by the requirement 
of fitting these building blocks to a closed contour in a very compact and symmetric manner.

There are several interesting topics for further studies. One should look for a geometrical pattern behind the BPS loops with vanishing
radiative corrections and its extension to higher orders. Combining the summation technique of \cite{Cavaglia:2018lxi} with our maps to conformal frames one should be able to generalise their results to non-planar circular triangles.  To supplement our weak coupling results  via AdS/CFT by a strong coupling analysis, one could try to relate
the conformal parameters $\alpha_j,\beta_j$ to the local conformal invariant functions along smooth contours used in \cite{Cairns},\cite{He:2017cwd} by promoting them to distributions.  On pure geometrical
level it would be nice to extend the closing conditions to higher circular polygons.\\[20mm]
%%%%%%%%%%%%%%%%%%%%%%%%%
{\bf Acknowledgement:}\\
I thank the Quantum Field and String Theory group at Humboldt University for kind virtual hospitality.\\[10mm]
%%%%%%%%%%%%%%%%%%%%%%
%%%%%%%%%%%%%%%%%%%%%%%%
\section*{Appendix A: Conformal geometry of circular triangles}\label{A}
This appendix is an extension of appendix D in \cite{Dorn:2020meb}. It handles the full 4D case and gives a nice symmetric form of the closing
condition in 3D, which was lacking in that paper. 

After a suitable translation,  we can map the corner number 3 by a conformal inversion to infinity and scale, with a subsequent
dilatation, the distance between the corners number 1 and 2 to one. Then by further using isometries we end up with a situation as follows.
The images of the corners are $Y_1=(0,0,0,0),~Y_2=(1,0,0,0),~Y_3=\infty$ and the circular edge number 3 is located in the $(1,2)$-plane and has negative $2$-coordinates. The circumcircle
is now the straight line along the $1$-axis and the images of the edges number 1 and 2 are parts of straight lines through $Y_2$ and $Y_1$, respectively.
The unit vectors  at $Y_1$ or $Y_2$  pointing in the direction of $Y_3$ along edges number 2 or 1 are  
\beq
e_j=(\sin\psi_j\sin\vartheta_j\cos\varphi_j,\,\sin\psi_j\sin\vartheta_j\sin\varphi_j,\,\sin\psi_j\cos\vartheta_j,\,\cos\psi_j)~.\label{ej}
\eeq
Since at each corner the tangents to the circumcircle and the adjacent edges span at most a three-dimensional subspace, we can choose e.g. $\psi_2=\frac{\pi}{2}$.
This fixes our conformal frame, and the following 6 conformal invariants characterise a circular triangle in 4D 
$$\psi_1,\vartheta_1,\vartheta_2,\varphi_1,\varphi_2,~~~\mbox{and}~~\beta_3~.$$
For an illustration of its 3D projection see figure 4 in  \cite{Dorn:2020meb}.
The three cusp and the two remaining torsion angles are then given by
\bea 
\cos\beta_1&=&\sin\vartheta_2\,\cos\varphi_2~,\label{beta-theta-phi-1}\\
\cos\beta_2&=&-\sin\psi_1\,\sin\vartheta_1\,\cos\varphi_1~.\label{beta-theta-phi-2}
\eea
\bea
\cos\alpha_1&=&\sin\psi_1\,\sin\vartheta_1\,\cos(\beta_3+\varphi_1)~,\label{alpha-theta-phi-1}\\
\cos\alpha_2&=&-\sin\vartheta_2\,\cos(\beta_3-\varphi_2)~,\label{alpha-theta-phi-2}\\
\cos\alpha_3&=&\sin\psi_1\,\big (\cos\vartheta_1\,\cos\vartheta_2+\sin\vartheta_1\,\sin\vartheta_2\,\cos(\varphi_1-\varphi_2)\big)\nonumber\\
&=&\sin\psi_1\,\big (\cos\vartheta_1\,\cos\vartheta_2+\sin\vartheta_1\,\sin\vartheta_2\,\sin\varphi_1\,\sin\varphi_2\big )-\cos\beta_1\,\cos\beta_2~.\label{alpha-theta-phi-3}
\eea
Using \eqref{beta-theta-phi-2} and  \eqref{alpha-theta-phi-1} to express $\vartheta_1,\varphi_1$ in terms of $\alpha_1,\beta_2,\beta_3$ we get
\bea
\sin^2\psi_1\,\sin^2\vartheta_1&=&\frac{\cos^2\alpha_1+\cos^2\beta_2+2\,\cos\alpha_1\,\cos\beta_2\,\cos\beta_3}{\sin^2\beta_3}~,\\
\cos^2\varphi_1&=&\frac{\cos^2\beta_2\,\sin^2\beta_3}{\cos^2\alpha_1+\cos^2\beta_2+2\,\cos\alpha_1\,\cos\beta_2\,\cos\beta_3}~,
\eea
and similarly with  \eqref{beta-theta-phi-1} and  \eqref{alpha-theta-phi-2}
\bea
\sin^2\vartheta_2&=&\frac{\cos^2\alpha_2+\cos^2\beta_1+2\,\cos\alpha_2\,\cos\beta_1\,\cos\beta_3}{\sin^2\beta_3}~,\\
\cos^2\varphi_2&=&\frac{\cos^2\beta_1\,\sin^2\beta_3}{\cos^2\alpha_2+\cos^2\beta_1+2\,\cos\alpha_2\,\cos\beta_1\,\cos\beta_3}~.
\eea
Inserting this in  \eqref{alpha-theta-phi-3}  we get\footnote{Taking into account, that all angles except the $\varphi$'s are in $(0,\pi)$ and that for $\varphi_j \in (0,2\pi )$ the signs of $\sin\varphi_j$  are correlated to that of  $q_j$ by
$\mbox{sign}\,q_j=-\mbox{sign}\,\sin\varphi_j$ for {\it both} $j=1$ and $j=2$.}
\beq
\sin\psi_1\;\cos\vartheta_1\,\cos\vartheta_2~=~(q_3-q_1q_2)\,\sin\beta_1\,\sin\beta_2\label{closing-zero}
\eeq
%%%%%%%%%%%%%%%%%%%%%%%%%%%%%%%%%%%%,
%% 
%% See notes page 66c ff , for sign of q_1,q_2 see 66l and 66m
%% 
%%%%%%%%%%%%%%%%%%%%%%%%%%%%%%%%%%%%%%%%%%%
with 
\beq
q_j~=~\frac{\cos\alpha_j+\cos\beta_{j+1}\,\cos\beta_{j-1}}{\sin\beta_{j+1}\,\sin\beta_{j-1}}~.\label{qj}
\eeq
After squaring \eqref{closing-zero} and a little bit more algebra we arrive at
\beq
1+2\,q_1\,q_2\,q_3-q_1^2-q_2^2-q_3^2\,=\,\frac{\cos^2\psi_1\cos^2\vartheta_2}{\sin^2\beta_1\sin^2\beta_2}~.\label{closing-one}
\eeq
Each $q_j$ is a parameter characterising a corner of the triangle with its two adjacent circular edges. For a planar circular triangle, or one which is completely located at a sphere, one has, e.g. for $q_3$, either
\beq
\alpha_3~=~\vert\pi-(\beta_1+\beta_2)\vert,~~~\mbox{\small or}~~~
\alpha_3~=~\pi-\vert \beta_1-\beta_2\vert~.\label{planar}
\eeq
This implies
\beq
q_j~=~\pm 1~~~~~\mbox{\small (planar case)}~,\label{q-planar}
\eeq
depending on whether the two adjacent edges are on the same or opposite sides of the circumcircle.\footnote{Note that in the planar case with  circular edges on the same side of the circumcircle for $\beta_1+\beta_2>\pi $ one has  to handle an additional UV divergence due to a crossing of the edges.}

In the generic 4D  situation  we have, 
\beq
\cos\alpha_3~=~-\cos\beta_1\cos\beta_2~+~\sin\beta_1\sin\beta_2\,\vec u_1\vec u_2~,
\eeq
with $\vec u_1$ and $\vec u_2$ three-dimensional unit vectors describing at corner number 3 the projection of the tangents  to the edges on the  subspace perpendicular to the tangent to the  circumcircle.
Hence instead of \eqref{planar} one gets,
\beq
\vert \pi-(\beta_1+\beta_2)\vert~\leq~\alpha_3~\leq~\pi-\vert\beta_1-\beta_2\vert\label{angles-generic}
\eeq
and cyclic permutations for the other two corners. As a consequence of the last inequalities we get
\beq
\vert q_j\vert~\leq ~1~. \label{q-pm1}
\eeq
Hence the $q_j$'s are some kind of off-planarity parameters for their corresponding corners with their two adjacent edges.  Varying a certain $q_j$ from $1$ to $-1$ one interpolates between the planar same side situation and the planar opposite side situation via additional  dimensions. 

Comparing $q_j$ with a standard measure for off-planarity, the three-dimensional Gram determinant $G_j$ of the tangent vectors at $X_j$ on the circumcircle and on the two neighbouring edges, we find
\beq
G_j~=~\sin^2\beta_{j-1}\sin^2\beta_{j+1}(1-q_j^2)~.
\eeq

We now come back to \eqref{closing-one}. With \eqref{beta-theta-phi-1} and \eqref{beta-theta-phi-2} one can bring its r.h.s. in the form
$$\frac{\cos^2\vartheta_2}{\cos^2\vartheta_2+\sin^2\vartheta_2\sin^2\varphi_2}\cdot \frac{\cos^2\psi_1}{\cos^2\psi_1+\sin^2\psi_1(\cos^2\vartheta_1+\sin^2\vartheta_1\sin^2\varphi_1)}~.$$
Obviously it is a positive number between zero and one and it depends on $\psi_1$, a parameter not correlated with the $q_j$'s, i.e.
\beq
1+2\,q_1\,q_2\,q_3-q_1^2-q_2^2-q_3^2\,\in\,(0,1)~.\label{q-interval}
\eeq
The off-planarity parameters $q_j$
characterise the building blocks, consisting of corners with adjacent edges. $\psi_1 $ parameterises the torsion between the three-dimensional subspaces spanned by
two of these building blocks.  

Then altogether we can state: The off-planarity parameters $q_j$ for the three building blocks are by their intrinsic geometry constrained by \eqref{q-pm1}. If one wants to combine them to a closed triangle in 4D, they have in addition to obey the inequality \footnote{The other bound from \eqref{q-interval} is automatically realised due to \eqref{q-pm1}.}\footnote{As an interesting side remark, note that the condition \eqref{closing-final} after $q_j=2u_j-1$ coincides with that for the allowed region of  the three cross ratios $u_1,u_2,u_3$, describing the conformal geometry of null hexagons \cite{Alday:2009dv,Dorn:2012cn }.}
\beq
1+2\,q_1\,q_2\,q_3-q_1^2-q_2^2-q_3^2\,\geq\,0~.\label{closing-final}
\eeq 
In 3D, i.e. $\psi_1=\frac{\pi}{2}$, the inequality has to be saturated. This then is related to the reduction of the number of conformal invariants from $6$ to $5$. In 2D, due to
\eqref{q-planar}, remains the condition $q_1q_2q_3=1$. It simply states, that only zero or two corners with opposite side edges are allowed for closing.

\begin{figure}[h]
\begin{minipage}{0.4\textwidth}
\includegraphics[width=\textwidth]{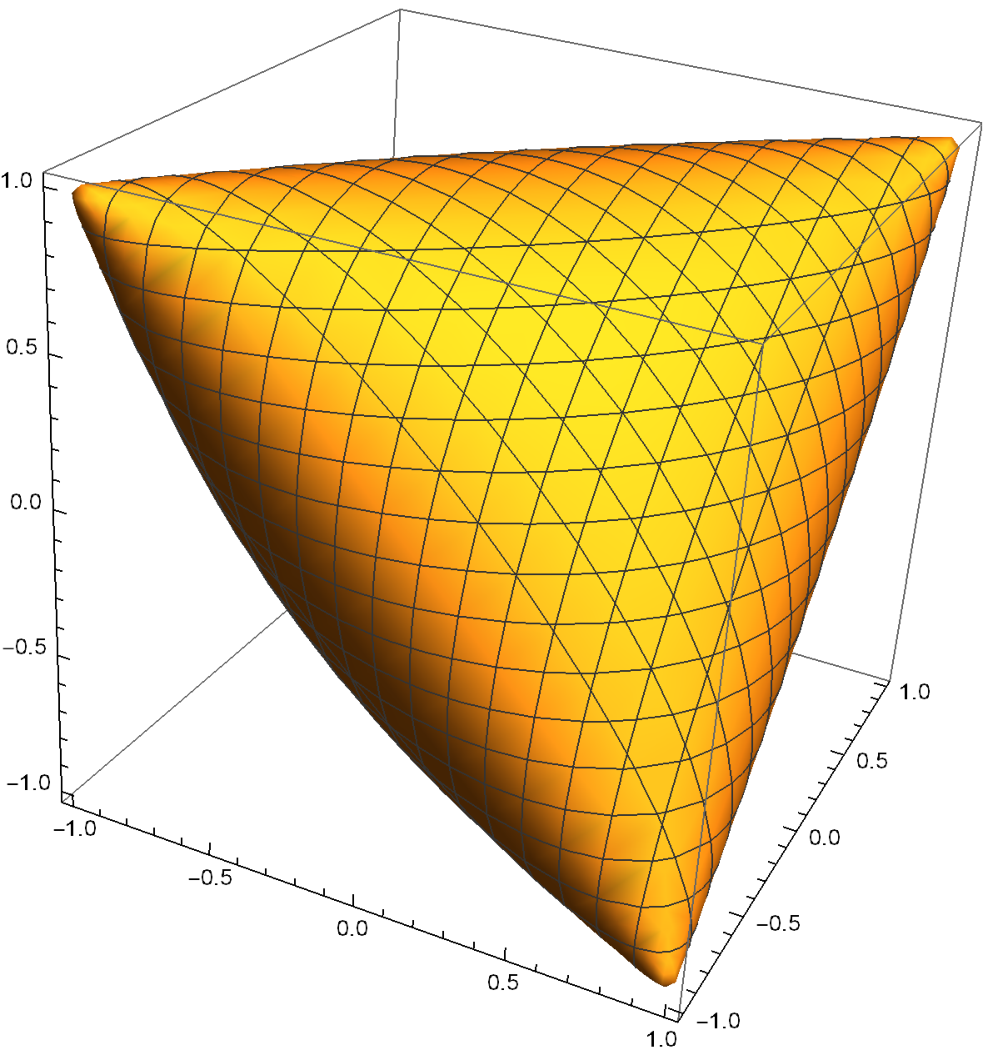}
\end{minipage}
\begin{minipage}{0.5\textwidth}
\includegraphics[width=\textwidth]{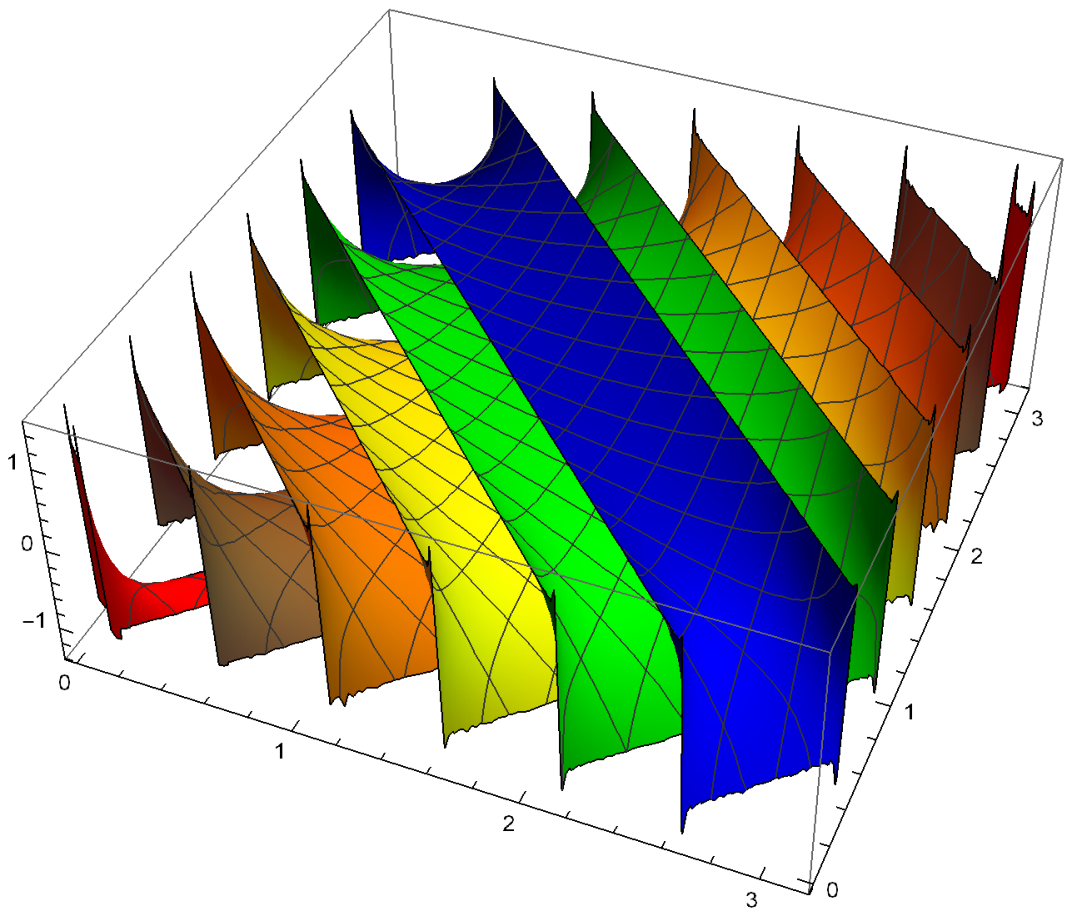}
\end{minipage}
\caption {\it Left: Locus of points $(q_1,q_2,q_3)$, allowed by the constraints \eqref{closing-final} and \eqref{q-pm1}. \\Right: \it $q_3$ as a function of $\beta_1$ and $\beta_2$ for various values of $\alpha_3$. Shown are equidistant steps for $\alpha_3$, starting from $0.5$ in blue up to $3$ in red.}
\label{q-plot}
\end{figure}

The points in $q$-space, constrained by  \eqref{q-pm1} and  \eqref{closing-final}  are illustrated in the left of fig.\ref{q-plot}.  For a circular triangle in 4D  one is restricted to the interior of  the rounded tetraeder, in 3D to its boundary and in 2D to one of its  4 corners.

For a fixed value of $\alpha_3$, the region allowed by \eqref{angles-generic} is a rectangle within the square $\beta_1,\beta_2\in (0,\pi)$, placed symmetrically
around the diagonals 
\beq
\pi-\alpha\leq\beta_1+\beta_2\leq\pi+\alpha,~~~\vert \beta_1-\beta_2\vert\leq \pi-\alpha~. \label{b1b2}
\eeq
Since $\beta_1+\beta_2$ on its upper bound is larger than $\pi$ this corresponds to the same side planar case with crossing edges. On the lower bound there is no crossing.

The function $q_3(\beta_1,\beta_2,\alpha_3)$ is visualised in the right part of fig.\ref{q-plot}. The boundaries of the rectangles correspond to the planar
situation. Along the lower and upper boundary for $\beta_1+\beta_2$ one has $q_3=1$ and on the lower and upper boundary for $\beta_1-\beta_2$ one has $q_3=-1$.\footnote{Since the limit for $q_3$ at the corners depends on the direction of its approach, the numerical evaluation near the corners becomes unstable. This explains the apparent spikes at the corners in fig.\ref{q-plot}.}     
           
%%%%%%%%%%%%%%%%%%%%%%%%%%%,%%i
 %%%%%%%%%%%%%%%%%%%%%%%%%%%%%%%%%%%%%%%%%%%%
%%%%%%%%%%%%%%%%%%%%%%%%%%%%%%%%%%%
%%%%%%%%%%%%%%%%%%%%%%%%%%%%%%%%
 \section*{Appendix B: Integral for scalar corner term }\label{J-calc}

Here we study the $b\rightarrow 0$ limit of
\beq
J(\alpha,\beta_1,\beta_2,b)=\int_0^{\infty}\frac{dt_1dt_2}{1+t_1^2+t_2^2-2t_1t_2\cos\alpha+2t_1\cos\beta_1+2t_2\cos\beta_2+b^2f(t_1,t_2)}~,
\label{app-I}
\eeq
where $f(t_1,t_2)$ has the structure
\beq 
f(t_1,t_2)~=~t_1^2t_2^2+g(t_1,t_2)~,
\eeq
with $g(t_1,t_2)$ a polynomial in $t_1$ and $t_2$ whose  overall degree is only three. It arises if \eqref{y-conf-frame} is inserted in \eqref{x-y}.

We are interested in the logarithmic divergence and the finite term for $b\rightarrow 0$ and argue at first, that for this purpose $g(t_1,t_2) $ can be neglected.
Denoting the denominator in\eqref{app-I} without the $b^2f$-term as $D$ we can write
\beq
J=\int_0^{\infty}\frac{dt_1dt_2}{D+b^2t_1^2t_2^2}+\int_0^{\infty}\frac{dt_1dt_2}{D+b^2t_1^2t_2^2}\Big (\big(1+\frac{b^2g(t_1,t_2)}{D+b^2t_1^2t_2^2}\big)^{-1}-1\Big )~.\label{app-I-split1}
\eeq
The maximum of $\vert \frac{b^2g(t_1,t_2)}{D+b^2t_1^2t_2^2}\vert  $ in the whole integration region is  of order ${\cal O}(b)$, hence we can conclude that the second
integral in \eqref{app-I-split1} is of order ${\cal O}(b\, \log b)$.

Then, after performing in the first integral the $t_2$-integration we get
\footnote{Strictly speaking the $t_2$-integration yields the expression, where $b^2(t^4+2t^3\cos\beta_1+t^2)$ stands instead of only $b^2t^4$. But repeating the argumentation  from just above, the error is again vanishing for $b\rightarrow 0$.}
\beq
J(\alpha,\beta_1,\beta_2,b)=\int_0^{\infty}
\frac{dt}{\sqrt{S(t)+b^2t^4}}
\Big(\frac{\pi}{2}-
\arctan\big(\frac{\cos\beta_2-t\,\cos\alpha}{\sqrt{S(t)+b^2t^4}}\big)\Big )~+~{\cal O}(b\,\log b)~,
\eeq
with
\beq
S(t)~=~t^2\,\sin^2\alpha+2t(\cos\beta_1+\cos\beta_2\,\cos\alpha)+\sin^2\beta_2 ~.\label{S}
\eeq
For $b=0$, the integrand behaves for $t\rightarrow\infty$ as $(\pi-\alpha)/(t\,\sin\alpha)$.  Trivially, this is the source for a logarithmic divergence. A little bit more effort
is needed to extract also the not divergent, but finite part in the limit $b\rightarrow 0$. 

To this end we split the integral  into pieces in a sequence of steps. First into
integrals from zero to one and from one to infinity. In the first piece $b$ can be put  to zero under the integral. Hence
\beq
J(\alpha,\beta_1,\beta_2,b)~=~J_0(\alpha,\beta_1,\beta_2)~+~J_1(\alpha,\beta_1,\beta_2,b)~+~{\cal O}(b\,\log b)+{\cal O}(b^2)~,\label{I0I1}
\eeq
with
\bea
J_0(\alpha,\beta_1,\beta_2)&=&\int_0^{1}
\frac{dt}{\sqrt{S(t)}}
\Big(\frac{\pi}{2}-
\arctan\big(\frac{\cos\beta_2-t\,\cos\alpha}{\sqrt{S(t)}}\big)\Big )~,\label{I0}\\
J_1(\alpha,\beta_1,\beta_2,b)&=&
\int_1^{\infty}
\frac{dt}{\sqrt{S(t)+b^2t^4}}
\Big(\frac{\pi}{2}-
\arctan\big(\frac{\cos\beta_2-t\,\cos\alpha}{\sqrt{S(t)+b^2t^4}}\big)\Big )~.
\eea
For the second split we concentrate on the first factor of the integrand in $J_1$, write\\[2mm] it as $1/\sqrt{t^2\sin^2\alpha+b^2t^4}+\big (1/\sqrt{S(t)+b^2t^4}-1/\sqrt{t^2\sin^2\alpha+b^2t^4}\big )$ and send  $b$ to zero in the subtracted term. Then we get
\bea
J_1(\alpha,\beta_1,\beta_2,b)&=&J_{11}(\alpha,\beta_1,\beta_2,b)~+~J_{10}(\alpha,\beta_1,\beta_2)~+~{\cal O}(b^2)~,\label{I10I11}\\[2mm]
J_{11}(\alpha,\beta_1,\beta_2,b)&=&\int_1^{\infty}
\frac{dt}{\sqrt{t^2\sin^2\alpha+b^2t^4}}\,
\Big(\frac{\pi}{2}-
\arctan\big(\frac{\cos\beta_2-t\,\cos\alpha}{\sqrt{S(t)+b^2t^4}}\big)\Big )~,\
\eea
\bea
J_{10}(\alpha,\beta_1,\beta_2)
~=~\int_1^{\infty}dt
\Big(\frac{1}{\sqrt{S(t)}}-\frac{1}{t\,\sin\alpha}\Big)
\Big(\frac{\pi}{2}-
\arctan\big(\frac{\cos\beta_2-t\,\cos\alpha}{\sqrt{S(t)}}\big)\Big )
.\label{I10}
\eea
So far $J_0$ and $J_{10}$ are finite and $J_{11}$ still divergent for $b\rightarrow 0$. The last splitting now concerns  $J_{11}$. Keeping in mind
$\arctan(-\cot\alpha))=\alpha-\pi/2$, we write
\bea
J_{11}(\alpha,\beta_1,\beta_2,b)&=&J_{11}^{(1)}(\alpha,b)+J_{11}^{(2)}(\alpha,\beta_1,\beta_2)+J_{11}^{(3)}(\alpha,\beta_1,\beta_2,b)+{\cal O}(b^2)~,
\label{I11}\\[2mm]
J_{11}^{(1)}(\alpha,b)&=&(\pi-\alpha)\int_1^{\infty}\frac{dt}{\sqrt{t^2\sin^2\alpha+b^2t^4}}~\label{I111},\\[2mm]
J_{11}^{(2)}(\alpha,\beta_1,\beta_2)&=&\int_1^{\infty}\frac{dt}{t\,\sin\alpha}\Big (\alpha-\frac{\pi}{2}-\arctan\big(\frac{\cos\beta_2-t\,\cos\alpha}{\sqrt{S(t)}}\big )\Big )~,\label{I112}
\\[2mm]
J_{11}^{(3)}(\alpha,\beta_1,\beta_2,b)&=&\int_1^{\infty}\frac{dt}{\sqrt{t^2\sin^2\alpha+b^2t^4}}\\[2mm]
&&~~~~~~\Big(\arctan\big(\frac{\cos\beta_2-t\,\cos\alpha}{\sqrt{S(t)}}\big )-\arctan\big(\frac{\cos\beta_2-t\,\cos\alpha}{\sqrt{S(t)+b^2t^4}}\big)\Big )~.\nonumber
\eea
$J_{11}^{(1]}$ is a standard integral and expressed as
\beq
J_{11}^{(1)}(\alpha,b)~=~\frac{\pi-\alpha}{\sin\alpha}\,\log\frac{2\,\sin\alpha}{b}~+~{\cal O}(b^2)~.
\eeq
To get under control the $b\rightarrow 0$ limit of $J_{11}^{(3)}$, we substitute the integration variable via $t=1/(bu)$ and arrive with \eqref{S} at
\bea
J_{11}^{(3)}(\alpha,\beta_1,\beta_2,b)&=&\int_0^{1/b}\frac{du}{\sqrt{1+u^2\sin^2\alpha}}\\[2mm]
&&\Big(\arctan\big(\frac{bu\,\cos\beta_2-\cos\alpha}{\sqrt{\sin^2\alpha+2bu(\cos\beta_1+\cos\beta_2\cos\alpha)+b^2u^2\sin^2\beta_2}}\big )\nonumber\\[2mm]
&&-\arctan\big(\frac{bu\,\cos\beta_2-\cos\alpha}{\sqrt{\sin^2\alpha+2bu(\cos\beta_1+\cos\beta_2\cos\alpha)+b^2u^2\sin^2\beta_2+1/u^2}}\big )\Big )
\nonumber
\eea
Now we see that the limit $b\rightarrow 0$ exists and find
\bea
J_{11}^{(3)}(\alpha,\beta_1,\beta_2,b)&=&\int_0^{\infty}\frac{du}{\sqrt{1+u^2\sin^2\alpha}}\label{I113}\\[2mm]
&&~~~~~~~~~~~~~\Big(\alpha -\frac{\pi}{2}+\arctan\big(\frac{u\,\cos\alpha}{\sqrt{1+u^2\sin^2\alpha}}\big )\Big )~+~{\cal O}(b)~.
\nonumber
\eea
Finally, collecting \eqref{I0I1},\eqref{I0},\eqref{I10I11},\eqref{I10}\eqref{I11},\eqref{I111},\eqref{I112},\eqref{I113} we get
\beq
J(\alpha,\beta_1,\beta_2,b)~=~{\cal O}(b\,\log b)~+~{\cal O}(b)~+~\frac{\pi-\alpha}{\sin\alpha}\,\log\frac{2\,\sin\alpha}{b}~+~Q(\alpha,\beta_1,\beta_2)~,\label{J-final}
\eeq
\bea
Q(\alpha,\beta_1,\beta_2)&=&\int_0^{\infty}\frac{dt}{\sqrt{1+t^2\sin^2\alpha}}\Big(\alpha -\frac{\pi}{2}+\arctan\big(\frac{t\,\cos\alpha}{\sqrt{1+t^2\sin^2\alpha}}\big )\Big )\label{Q}
\\[2mm]
&&+~\int_0^1\frac{dt}{\sqrt{S(t)}}\,\Big (\frac{\pi}{2}-\arctan\big (\frac{\cos\beta_2-t\,\cos\alpha}{\sqrt{S(t)}}\big )\Big )\nonumber\\[2mm]
&&+~\int_1^{\infty}dt\,\left (\frac{\alpha-\pi}{t\,\sin\alpha}~+~\frac{1}{\sqrt{S(t)}}\Big (\frac{\pi}{2}-\arctan\big (\frac{\cos\beta_2-t\,\cos\alpha}{\sqrt{S(t)}}\big )\Big )\right )~,\nonumber
\eea
with $S$ as function of $\alpha,\beta_1,\beta_2,t$ defined in  \eqref{S}.

For the first integral in \eqref{Q}, let us call it $Q_0(\alpha)$, we found a representation in terms of log's and dilog's
\beq
Q_0=\frac{1}{\sin\alpha}\left ( \big (\alpha-\pi\,\Theta(\alpha-\frac{\pi}{2})\big )\log\,\sin\alpha -\frac{1}{2}\mbox{Im}\,\mbox{Li}_2(-e^{2i\alpha})+\mbox{Im}\,\mbox{Li}_2\big (\frac{1-e^{2i\alpha}}{2}\big )\right ). \label{Q0}
\eeq
It remains an interesting open question, whether there is a similar representation also for the other two integrals, which depend besides on $\alpha$ also on  the two torsion angles.  In the simpler planar case  $S(t)$ is a pure square, and we present such a representation of the corresponding  full $Q$  in appendix  D, see \eqref{Qsame},\eqref{Qoppo}.
%%%%%%%%%%%%%%%%%%
%%
%%  direct numerical check of the last  formula as typed in this file: See 2020-05-28.nb
%%
%%%%%%%%%%%%%%%%%%%%%%%%%

For the case  $C_3$ in the main text take  $\alpha=\alpha_3,~~b=\frac{D_{12}}{D_{23}D_{13}} a$ and cyclic permutations for the other $C_j$.

Our function $Q(\alpha,\beta_1,\beta_2)$ was constructed as the finite part after subtracting the UV divergent terms due to the cusp at the corner under consideration. As stated in the previous appendix,  along the upper bound of $\beta_1+\beta_2$ one has a planar situation with crossing edges. This so far neglected additional
UV divergence has to show up as a divergence of $Q$. 

To illustrate this issue, let us start with the observation, that the only potential source  for a divergence of $Q$ could be related to zeros of the quadratic polynom $S(t)$. Inside \eqref{b1b2} they are either complex or negative,  hence outside the integration region for $t$. On the lower boundary of $\beta_1+\beta_2$ (same side planar case without crossing) one finds a double zero at negative values of $t$, i.e. no divergence. A double zero at positive $t$-values on gets on the
upper boundary of $\beta_1+\beta_2$ and on both boundaries of $\beta_1-\beta_2$.  At a zero $t_0$ of $S(t)$ the $\arctan$ in the second and third integral in  \eqref{Q} approaches   $\pm\frac{\pi}{2}$, depending on the sign of $\cos\beta_2-t_0\cos\alpha$. It turns out to be positive on  both boundaries
of $\beta_1-\beta_2$, the opposite side planar cases, and negative on the upper boundary of $\beta_1+\beta_2$. Therefore, $Q$ is only divergent along the upper boundary of $\beta_1+\beta_2$.

\begin{figure}[h]
\begin{minipage}{0.5\textwidth}
\includegraphics[width=\textwidth]{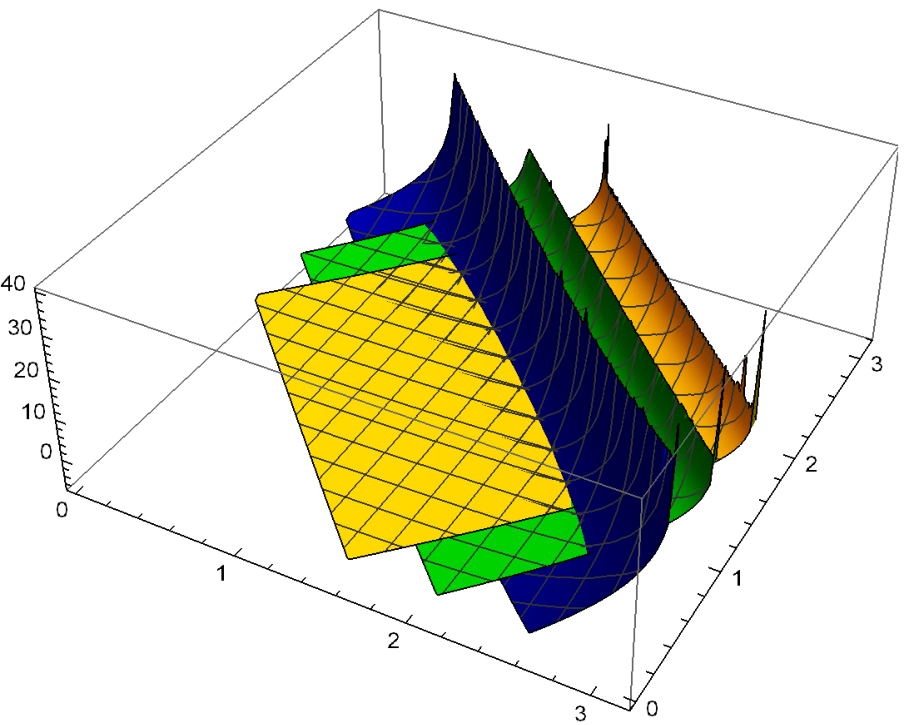}
\end{minipage}
\begin{minipage}{0.5\textwidth}
\includegraphics[width=\textwidth]{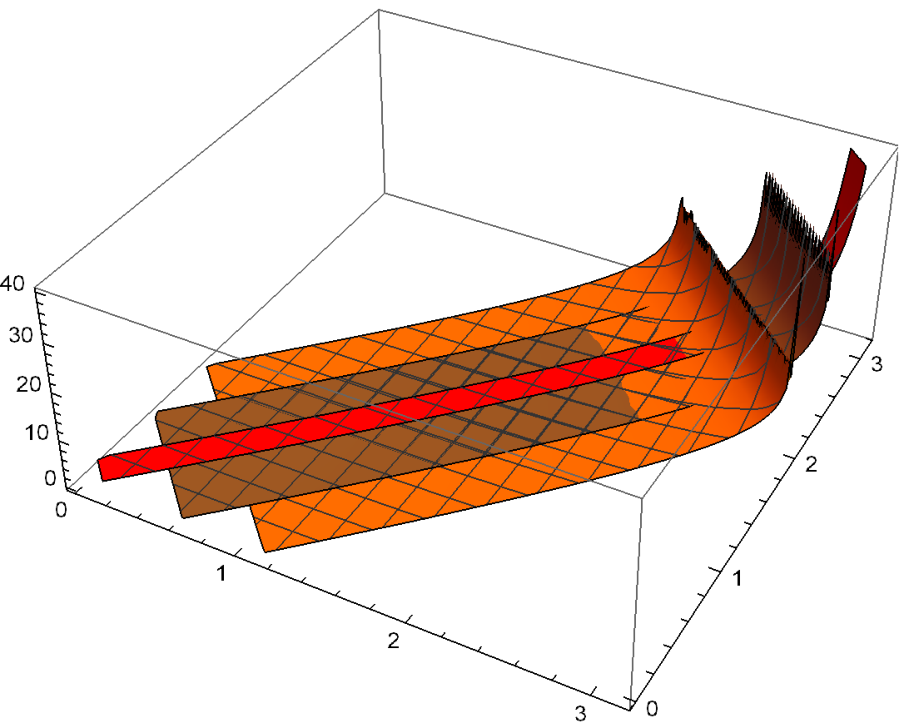}
\end{minipage}
\caption {\it The function $Q(\alpha,\beta_1,\beta_2)$ in dependence on $\beta_1,\beta_2$ for different values \\of $\alpha$.  On the left in blue, green, yellow for $\alpha=0.5,1.0,1.5$ and on the right in orange, brown and red for $\alpha =2.0,2.5,3.0$.}
\label{remainder3D}
\end{figure}

\begin{figure}[h]
\begin{minipage}{0.5\textwidth}
\includegraphics[width=\textwidth]{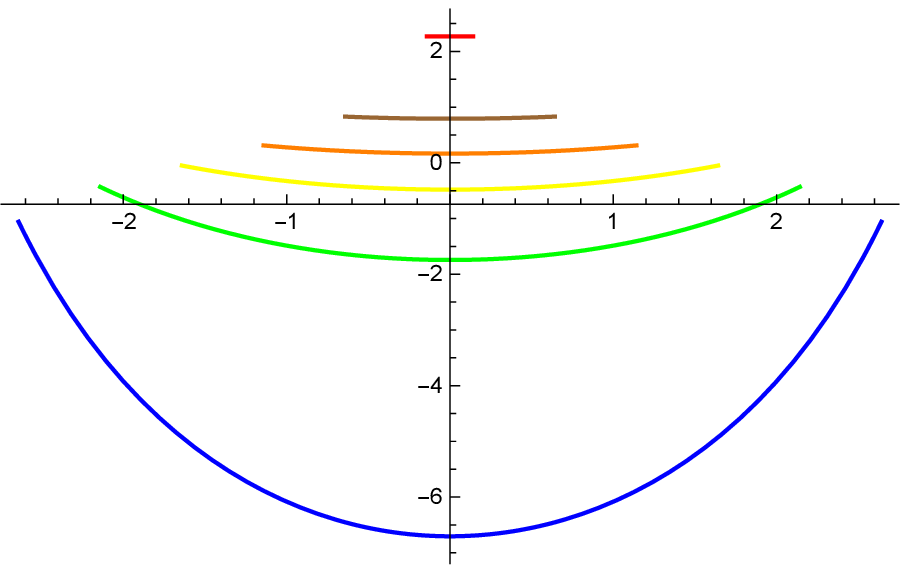}
\end{minipage}
\begin{minipage}{0.5\textwidth}
\includegraphics[width=\textwidth]{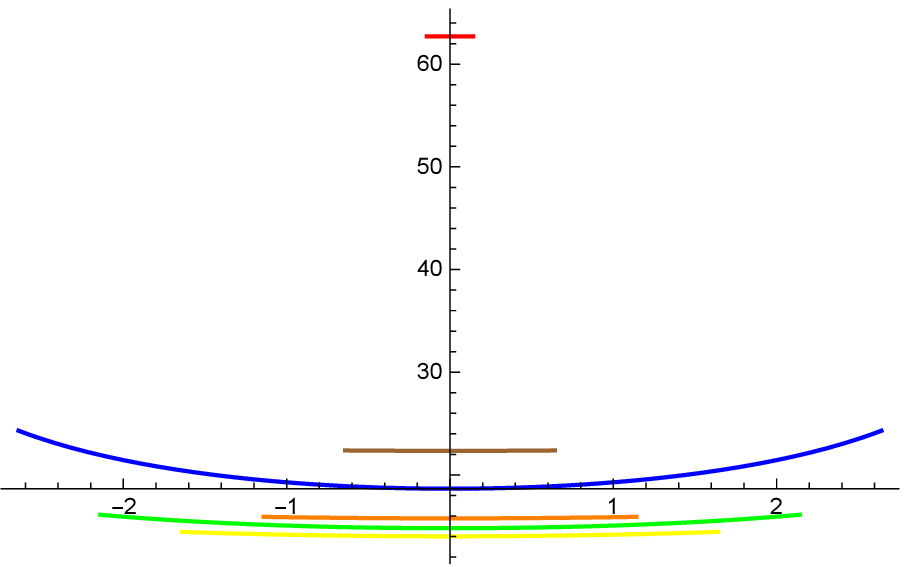}
\end{minipage}
\caption {\it $Q$  as a function of $\beta_1-\beta_2$. On the left at the lower boundary of $\beta_1+\beta_2$ and on the right at a distance 0.02  from the upper boundary of $\beta_1+\beta_2$.  The corresponding values of $\alpha$ are encoded in colours as in fig.\ref{remainder3D}.
}
\label{remainder3D-cuts}
\end{figure}

We present some plots of $Q$ in fig.\ref{remainder3D}, with a small neighbourhood of the  upper bound of $\beta_1+\beta_2$ excluded. A hasty view might suggest the impression that the shown bands would be flat. To avoid this mistake, we show cuts along
the lines of constant $\beta_1+\beta_2$  at the lower and very near to the upper end in fig.\ref{remainder3D-cuts}.
%%%%%%%%%%%%%%%%%%%%%%%%%%%%%%%%%%%%%%%%%%%%
\newpage
%%%%%%%%%%%%%%%%%%%%%%%%%%%%%%%%%%%
%%
%% Both figures generated  in 2020-10-22, There also some games concerning the divergent behaviour on upper bound
%% of b1+b2, Mathematica is there only restricted trustworthy, it says e.g. for NIntegrate[[1/Abs[x - 1], {x, 0, 2}] expect singularity, but gives nevertheless
%%  the number 158.752   see also older versions in  2020-05-19a.nb
%%
%%%%%%%%%%%%%%%%%%%%%%%%%%%%%%%%%%%
 %%%%%%%%%%%%%%%%%%%%%%%%%%%%%%%%%%%%%%%%%%%%
%%%%%%%%%%%%%%%%
%%%%%%%%%%%%%%%%%%%%%%%%%
\section*{Appendix C: Correction term for map to conformal frame }\label{DG}
In this appendix we sketch the calculation of the correction terms $A_j$ in \eqref{C3}, generated by the mapping to the respective conformal frame. We write the formulas for $A_3$, the other cases one gets by cyclic permutation of indices. 

Under a conformal inversion $x=\frac{w}{w^2}$ one has
\bea
(x_1-x_2)^2&=&\frac{1}{w_1^2w_2^2}(w_1-w_2)^2~,\nonumber\\
\dot x_j^{\mu}&=&\frac{1}{w_j^2}I^{\mu\nu}(w_j)\dot w_j^{\nu}~,~~~\mbox{with}~~~I^{\mu\nu}(w)=\delta^{\mu\nu}-2 \frac{w^{\mu}w^{\nu}}{w^2}~.
\eea
This guarantees the invariance of the unregularised scalar contribution used  in \eqref{x-y}, but implies for the vector contribution
\beq
\frac{\dot x_1\dot x_2}{(x_1-x_2)^2}~=~\frac{I^{\mu\lambda}(w_1)I^{\nu\lambda}(w_2)\dot w_1^{\mu}\dot w_2^{\nu}}{(w_1-w_2)^2}~,
\eeq
which leads to \cite{Drukker:2000rr}
\bea
\frac{\dot x_1\dot x_2}{(x_1-x_2)^2}~=~\frac{\dot w_1\dot w_2}{(w_1-w_2)^2}&+&\dot w_1^{\mu}\dot w_2^{\nu}\,\Big (\frac{w_2^{\nu}}{w_2^2}\,\partial_1^{\mu}\log (w_1-w_2)^2\\&+&\frac{w_1^{\mu}}{w_1^2}\,\partial_2^{\nu}\log (w_1-w_2)^2-\frac{1}{2}\partial_1^{\mu}\log w_1^2\,\partial_2^{\nu}\log w_2^2\Big )~.\nonumber \label{Drukker-Gross}
\eea
The derivative structure of the correction term leads to the invariance of the related integral for closed contours not passing the origin and to the anomaly
discussed in \cite{Drukker:2000rr},  if the origin is on the contour. But for our corner building block we have to handle an open (part of) contour.

At the beginning of  appendix A we have described a stepwise map of a corner building block, e.g. corner number 3 with adjacent edges,  to its conformal frame. Among these steps only the second one, an inversion, contributes to $A_3$. Taking into account the preceding translation to move corner number 3 to the origin, the images of the corners after the inversion are
\beq
W_1~=~\frac{X_1-X_3}{D_{13}^2}~,~~~W_2~=~\frac{X_2-X_3}{D_{23}^2}~,~~~W_3~=~\infty~.\label{XW}
\eeq 
The images of the edges in $w$-space are then parameterised
by
\beq 
w_1(t_1)~=~W_2+t_1\,n_2~,~~~w_2(t_2)~=~W_1+t_2\,n_1~,
\eeq
and in the sense of a universal contour parameter along the whole corner contribution we have
\beq
\dot w_1\,\dot w_2~=~-n_1n_2~=~-\cos\alpha_3~.
\eeq
The derivative structure of \eqref{Drukker-Gross} allows to perform trivial integrations, and we get with a preliminary IR cutoff $\Lambda$
\bea
A_3&=&\int_0^{\Lambda}\frac{dt\,(W_1n_1+t)}{t^2+2tW_1n_1+W_1^2}\big (\log (W_2-W_1+\Lambda n_2-tn_1)^2-\log (W_2-W_1-tn_1)^2\big)\nonumber\\[2mm]
&&+(1\leftrightarrow 2)-\frac{1}{2}\big (\log (W_1+\Lambda n_1)^2-\log (W_1^2)\big )\big (\log (W_2+\Lambda n_2)^2-\log (W_2^2)\big ).
\eea
Let us call the integral in the first line of the formula above $G_{13}$ (integration between corner 1 and corner 3) and expand the last term for large $\Lambda$ we get
\beq
A_3~=~G_{13}+G_{23}-2(\log\Lambda)^2+\,\log\Lambda\,(\log W_1^2+\log W_2^2)-\frac{\log W_1^2\,\log W_2^2}{2}~+~{\cal O}\big (\frac{\log\Lambda}{\Lambda}\big ).\label{A3} 
\eeq
Introducing the indefinite integral
\beq
H(K,L,M,N,x)~=~\int \frac{x+K}{x^2+2Kx+L}\,\log (x^2+2Mx+N)\,dx~,\label{H}
\eeq
we get
\bea
G_{13}&=&H(K,L,\hat M,\hat N,\Lambda)~+~H(K,L,\tilde M,\tilde N,\Lambda)\nonumber\\
&&-~H(K,L,\hat M,\hat N,0)~-~H(K,L,\tilde M,\tilde N,0)~,
\eea
with
\bea
K&=&W_1n_1~,~~~L~=~W_1^2~,~~~\hat M~=~(W_1-W_2)n_1~-~\Lambda\,\cos\,\alpha_3~,\\[2mm]
\hat N &=&(W_2-W_1)^2+2\Lambda(W_2-W_1)n_2+\Lambda ^2,~\tilde M=(W_1-W_2)n_1,~\tilde N~=(W_2-W_1)^2.\nonumber
\eea
Using the invariance of angles under the mapping and that the image of the circumcircle is given by the straight line passing $W_1$ and $W_2$, we can express the
just introduced constants via \eqref{XW} in terms of parameters of the original setting ($\delta_j$ defined in \eqref{delta-j}, $D_3$ in \eqref{b3})
\bea
K&=&\frac{1}{D_{13}}\,\cos\frac{\delta_2}{2}~,~~~L~=~\frac{1}{D_{13}^2}~,~~~\hat M~=~D_3\,\cos\beta_2-\Lambda\,\cos\,\alpha_3~,\nonumber\\[2mm]
\hat N&=&D_3^2+2\Lambda D_3\,\cos\beta_1+\Lambda^2~,~~~\tilde M~=~D_3\,\cos\beta_2~,~~~\tilde N~=~D_3^2~.
\eea

As defined in \eqref{H} $H$, is given by a  sum of logarithms and dilogarithms, in which some of the terms have  complex arguments. Neither for $G_{13}$, nor for the combination $ G_{13}+G_{23}$, needed for $A_3$ in \eqref{A3}, we could  eliminate the dilogarithms in favour of logarithms via their standard functional relations. However,
one can reach this goal for the sum of all $G_{ij}$ terms appearing in the sum $A_1+A_2+A_3$. Then after a massage of the many remaining logarithms, careful handling
of phases of complex terms and expanding for large IR cutoff $\Lambda$ we find cancellation of the IR divergent terms and finally a very compact expression
for the remaining finite contribution
\beq
A_1+A_2+A_3~=~\pi^2~+~\sum _j\big (\beta_j^2~-~(\pi-\alpha_j)^2~-~\frac{1}{4}\delta_j^2\big )~.\label{sum-A}
\eeq
%%%%%%%%%%%%%%%%
%%%%%%%%%%%%%%%%%%%%%%%%%
\section*{Appendix D: Function $Q$ in the planar case in terms of standard functions }
In this appendix we specialise in the planar case and  sketch the representation of the whole function $Q$, given in \eqref{Q} by three convergent one-dimensional integrals, in terms of standard functions.

The function $S(t)$, defined by \eqref{S}, can be rewritten as
\beq
S(t)=\sin^2\alpha \left(\Big(t+\frac{\cos\beta_1+\cos\beta_2\,\cos\alpha}{\sin^2\alpha}\Big)^2~+~\frac{(1-q^2)\,\sin^2\beta_1\sin^2\beta_2}{\sin^4\alpha}\right)~,
\eeq
with the off-planarity parameter $q$ defined according to \eqref{qj}.  In the planar case we have $\vert q\vert =1$ and $S$ becomes a pure square. Furthermore, the three angles $\alpha,\beta_1,\beta_2$ are no longer independent. With \eqref{planar} we get
\beq
S(t)\vert _{\mbox{\scriptsize planar}}~=~\big (t\,\sin\beta+\sin\beta_2\big)^2,~~~~~\mbox{with}~~~\beta = \beta_1+\beta_2~~\mbox{or}~~\beta=\vert\beta_1-\beta_2\vert~
\eeq
in the cases where the edges are on the same or opposite side of the circumcicle, respectively.

\begin{figure}[h!]
\begin{center}
 \includegraphics[width=15cm]{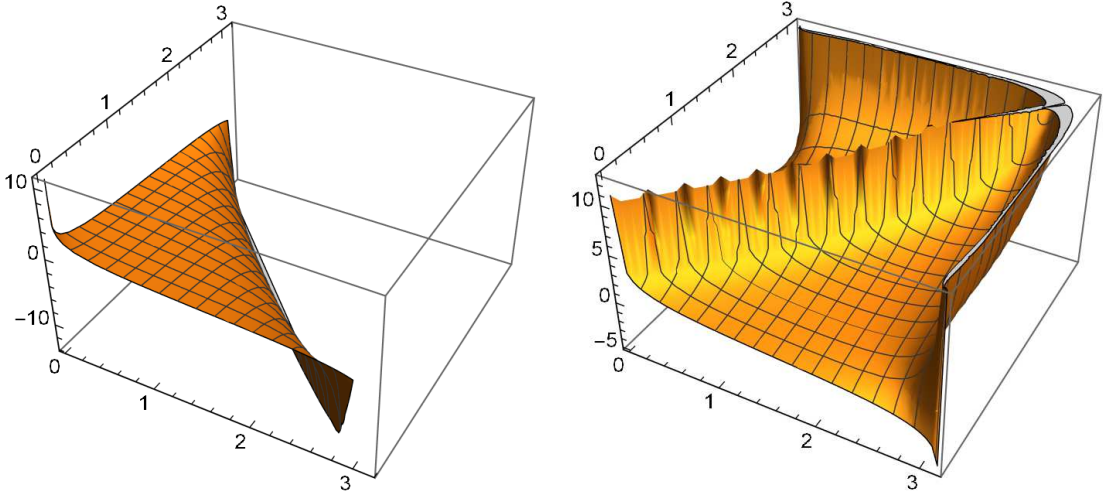}
\end{center}
\caption {\it On the left: $Q^{\mbox{\scriptsize planar}}_{\mbox{\scriptsize same}}(\beta_1,\beta_2)$ for $\beta_1+\beta_2<\pi$. On the right:  $Q^{\mbox{\scriptsize planar}}_{\mbox{\scriptsize oppo}}(\beta_1,\beta_2)$ for all $\beta_1,\beta_2\in(0,\pi).$ }
\label{same-oppo}
\end{figure}

Then the only nontrivial integrations, needed in the second and third row of \eqref{Q} are of the type
\bea
U(t_1,t_2)&=&\int_{t_1}^{t_2}\frac{d\,t}{t\,\sin\beta+\sin\beta_2}\,\arctan\Big(\frac{\cos\beta_2+t\,\cos\beta}{\sin\beta_2+t\,\sin\beta}\Big)\\[2mm]
&=&\frac{1}{\sin\beta}\,\int_{x(t_2)}^{x(t_1)}\frac{d\,x}{x}\,\arctan(\cot\beta+x),~~\mbox{\small with}~~x(t_j)=\frac{\sin(\beta-\beta_2)}{t_j\sin^2\beta+\sin\beta\sin\beta_2}~.\nonumber
\eea
The related indefinite integral is
\beq
\int \frac{d\,x}{x}\,\arctan(A+x)~=~\mbox{Im}\Big\{\,\log\big(-\frac{x}{A-i}\big)\,\log\big(1+i(A+x)\big)-\mbox{Li}_2\Big(\frac{x+A+i}{A+i}\Big)\Big\}~.
\label{K}
\eeq
Let us now remember, that due to their geometrical meaning $\beta_1,\beta_2\in (0,\pi)$. In the same side case we restrict ourselves to $\beta_1+\beta_2<\pi$, because otherwise the two edges intersect, causing 
divergences. In the opposite side case one always has $\beta<\pi$ and we assume in addition  w.l.o.g. $\beta_2>\beta_1$. Therefore the boundary values  needed for the $x$-integrations in \eqref{K} are positive in the same side case and negative in the opposite side case. With this in mind and taking into account  \eqref{Q0}, we finally get for
\bea
Q^{\mbox{\scriptsize planar}}_{\mbox{\scriptsize same}}(\beta_1,\beta_2)&=&Q\big(\pi-(\beta_1+\beta_2),\,\beta_1,\,\beta_2\big)~, ~~~~\beta_1+\beta_2<\pi\nonumber\\[2mm]
Q^{\mbox{\scriptsize planar}}_{\mbox{\scriptsize oppo}}(\beta_1,\beta_2)&=&Q\big(\pi-\vert\beta_1-\beta_2\vert,\,\beta_1,\,\beta_2\big)~,
\eea
the following expressions in terms of standard functions\footnote{The results, obtained by performing the integrations 
in the chosen order, confess the symmetry in the $\beta_j$  only after using some functional relations. For convenience we have written \eqref{Qsame} in an obviously symmetrised manner.}
\bea
Q^{\mbox{\scriptsize planar}}_{\mbox{\scriptsize same}}(\beta_1,\beta_2)&=&\frac{1}{2\,\sin(\beta_1+\beta_2)}\Big\{ \mbox{Im}\Big [ \mbox{Li}_2\Big(
\frac{\sin(\beta_1+\beta_2)}{\sin\beta_1}e^{-i\beta_2} \Big )+\mbox{Li}_2\Big(\frac{\sin(\beta_1+\beta_2)}{\sin\beta_2}e^{-i\beta_1} \Big )  \nonumber\\[2mm]
&&~~~~~~~~+\,2\,\mbox{Li}_2\Big(\frac{1-e^{-2i(\beta_1+\beta_2)}}{2}\Big)-\mbox{Li}_2\big(-e^{-2i(\beta_1+\beta_2)}\big )\Big ]\nonumber\\[2mm]
&&~~~~~~~~+\,(\beta_1-\beta_2)\,\log\,\frac{\sin\beta_2}{\sin\beta_1}+\pi\,\log\,\frac{\sin^2(\beta_1+\beta_2)}{\sin\beta_1\sin\beta_2}\nonumber\\[2mm]
&&~~~~~~~~+\,2\big( \pi\,\Theta(\beta_1+\beta_2-\frac{\pi}{2})-\beta_1-\beta_2\big)\,\log\,\sin(\beta_1+\beta_2)\Big\}~.\label{Qsame}
\eea
\bea
Q^{\mbox{\scriptsize planar}}_{\mbox{\scriptsize oppo}}(\beta_1,\beta_2)&=&\Big\{\mbox{Im\,Li}_2
\Big(\frac{\sin \vert\beta_1-\beta_2\vert} {\sin\,\mbox{max} (\beta_1,\beta_2)}
 e^{i\,\mbox{\scriptsize min} (\beta_1,\beta_2)}\Big )-\frac{1}{2}\mbox{Im\,Li}_2\Big(-e^{-2i\vert \beta_1-\beta_2\vert}\Big)
\nonumber\\[2mm]
&&+\,\mbox{Im\,Li}_2\Big(\frac{1-e^{-2i\vert\beta_1-\beta_2\vert}}{2}\Big)  +\mbox{min}(\beta_1,\beta_2)\,\log\,\frac{\sin\,\mbox{min}(\beta_1,\beta_2)}{\sin\,\mbox{max}(\beta_1,\beta_2)}\\[2mm]
&&+\,\big( \pi\,\Theta(\vert\beta_1-\beta_2\vert-\frac{\pi}{2})-\vert\beta_1-\beta_2\vert\big)\,\log\,\sin\vert\beta_1-\beta_2\vert\Big\}\frac{1}{\sin\vert \beta_1-\beta_2\vert}~.\label{Qoppo}\nonumber
\eea
Plots of these two functions are shown in fig.\ref{same-oppo}. Both show runaway behaviour at some locations. In the same side case it is related to the
approach to a situation with crossing edges at $\beta_1+\beta_2=\pi$, and at $\beta_1+\beta_2=0$ it is necessary to fit the smooth contour case ($\alpha=\pi$) as discussed around \eqref{circle-limit}. In the opposite side case the  last comment also explains  the singularity along the line $\beta_1=\beta_2$ . To understand the singularities  along $\beta_1=\pi$ and  $\beta_2=\pi$, one has to realise that if one of the  $\beta_j$ is equal to $\pi$, the corresponding edge is a piece of the circumcircle passing also the endpoint of the other edge. 
%%%%%%%%%%%%%%%%%%%%%%%%%%%%%%%%%%%%%%%%%%%%
%%%%%%%%%%%%%%%%%%%%%%%%%%%%%%%%%%%%%%%%%%%%%%%

For e.g. $0<\beta_2<\pi$ and $\beta_1\rightarrow 0$ the two planar versions of $Q$ should coincide, since this just corresponds to the smooth transition between
a same side and an opposite side situation.   Indeed one finds
\bea
Q^{\mbox{\scriptsize planar}}_{\mbox{\scriptsize same}}(0,\beta_2)&=&Q^{\mbox{\scriptsize planar}}_{\mbox{\scriptsize oppo}}(0,\beta_2)\\[2mm]
&=&\frac{1}{\sin\beta_2}\Big\{\mbox{Im\,Li}_2\Big(\frac{1-e^{-2i\beta_2}}{2}\Big) -\frac{1}{2}\mbox{Im\,Li}_2\Big(-e^{-2i\beta_2}\Big)\nonumber
\\[2mm]
&& ~~~~~~~~~~~~~~~~~~~~~~+\,\big(\pi\,\Theta(\beta_2-\frac{\pi}{2})-\beta_2\big)\,\log\,\sin\beta_2\Big\}~.\nonumber
\eea
 %%%%%%%%%%%%%%%%%%%%%
%%%%%%%%%%%%%%%%%%%%%%%%%

\end{document}